# 3D Theory of Microscopic Instabilities Driven by Space-Charge Forces

Vladimir Litvinenko*[1,2], Yichao Jing[2,1], Jun Ma[2], Irina Petrushina[1], Kai Shih[1], Gang Wang[2,1]

[1] Department of Physics and Astronomy, Stony Brook University, Stony Brook, NY
[2] Collider-Accelerator Department, Brookhaven National Laboratory, Upton, NY

*Abstract*

Microscopic, or short-wavelength, instabilities are known for drastic reduction of the beam quality and strong amplification of the noise in a beam. Space charge and coherent synchrotron radiation are known to be the leading causes for such instabilities. In this paper we present rigorous 3D theory of such instabilities driven by the space-charge forces. We define the condition when our theory is applicable for an arbitrary accelerator system with 3D coupling. Finally, we derive a linear integral equation describing such instability and identify conditions when it can be reduced to an ordinary second order differential equation.

## I. Introduction

High quality lepton and hadron beams play important role in various applications of accelerators ranging from colliders to X-ray free-electron lasers (FELs) [1-14]. These beams undergo through processes of generation, acceleration, transport, and compression, during which instabilities could cause significant degradation of beam's quality. On the other hand, some of these instabilities can be tamed and used for generation of coherent radiation [15-19] or hadron-beam cooling [20-23].

There are several 1D theories of micro-bunching instabilities [24-32], but none of them fully account for coupling between instabilities in transverse and longitudinal degrees of freedom. In this paper, we attempt to develop general 3D-theory of microscopic instabilities driven by space charge (SC) forces. There are compelling theoretical and experimental reasons why coupling between various degrees of motion should be included in the analysis of SC-driven instabilities. In fact, we observed a variety of coupled SC-driven instabilities in our superconducting accelerator and its beamlines [33]. Two measured beam profiles illustrating such coupling in SC driven instabilities are shown in Fig. 1: Fig. 1(a) is an example of strong coupling between radial and axial modes, while Fig. 1(b) exemplifies coupling between longitudinal and transverse (vertical) modes.

In this paper we are considering an accelerator with most general beam transport, described by a symplectic 6x6 transport map, which includes all macroscopic effects including SC forces. By linearizing the transport maps (in a vicinity of a chosen phase space trajectory) we reduce self-consistent Vlasov equation to a linear integral equation describing evolution of 3D Fourier harmonics of beam's density distribution.

For the derivations presented in this paper we pursue the classical plasma physics methods that are specifically modified for the modern accelerator lingo:

1. We consider an accelerator without any limitation on its components, acceleration, deceleration, compression, focusing, coupling, or its 3D beam trajectory.
2. We use the length along the reference trajectory, *s*, as an independent variable. Particle motion is described as evolution of a full set of 6 canonical variables driven by the Hamiltonian, which includes macroscopic SC forces.

3. We assume that effects of microscopic instability can be treated as a perturbation.
4. We consider that the beam transport map is evaluated as function of *s* for the unperturbed Hamiltonian including all macroscopic effects, and it is known from beam dynamics simulations.
5. We use Canonical transformation to the initial condition to remove macroscopic components and arrive to the linearized Vlasov equation.
6. We identify range when and where our microscopic approach is applicable and derive equation for perturbation Hamiltonian.
7. We use local linearization of the transport map with symplectic 6x6 matrix in Alex Dragt's notation [34]. The use of this notation allows to clearly identify the roles of the 3x3 matrix blocks in the evolution of the beam and the perturbation parameters.
8. We apply Fourier transform and arrive to explicit form of a linear integral equation describing evolution of the microscopic perturbations.
9. Finally, we identify conditions when the linear integral equation can be reduced to an ordinary second order differential equation for the electron beam density perturbation.

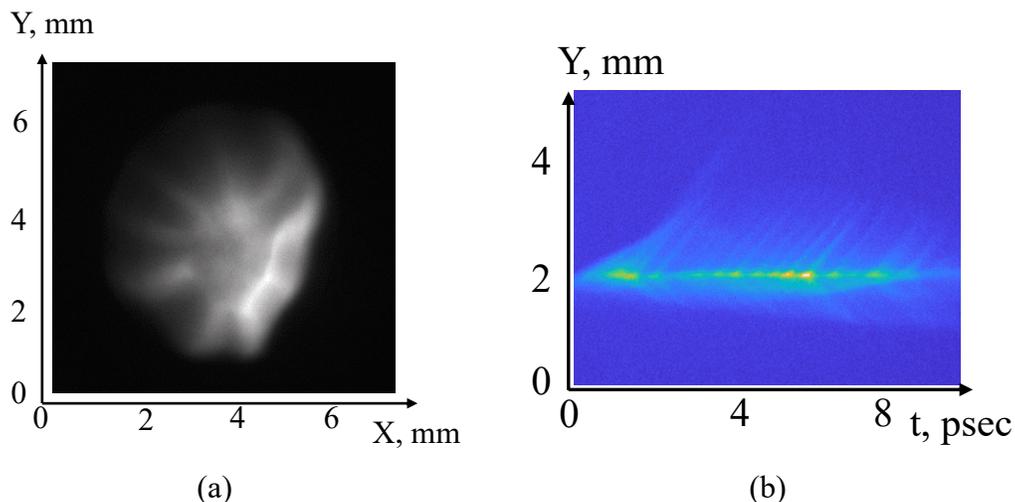

(a)        (b)

Figure 1. Samples of measured electron beam distributions in the CeC accelerator [33] illustrate some aspects of 3D coupling in SC-driven instabilities: (a) Coupling between radial and axial modes in SC-driven instability; (b) Feather-like coupling between vertical and longitudinal degrees of freedom in SC-driven instability.

It was very tempting to expand our approach to include coherent synchrotron radiation (CSR) effects. But such inclusion requires non-local interactions, which is outside of the scope of this paper. The CSR inclusion would, at least, double the length of this paper and make it convoluted.

To keep the main portion of the text compact, we appended the detailed discussions and derivation in five Appendices.

## II. Theory

Let us consider a charged particle beam with a reference particle moving along a curved trajectory $\vec{r}_o(s)$. Its motion can be described using a standard Frenet-Serret coordinate system with three unit orthogonal vectors $(\hat{e}_1, \hat{e}_2, \hat{e}_3)$ and the length along the reference trajectory $s = \int |d\vec{r}_o|$ (azimuth) serving as an independent variable [34-36]:

$$\hat{e}_3(s) = \frac{d\vec{r}_o}{ds};\ \hat{e}_1(s) = -\frac{\vec{r}_o''}{|\vec{r}_o''|};\ \hat{e}_2(s) = [\hat{e}_1(s) \times \hat{e}_3(s)];$$

$$\hat{e}_3' = -K_o(s) \cdot \hat{e}_1(s);\ \hat{e}_1' = K_o(s) \cdot \hat{e}_3(s) - \kappa_o(s) \cdot \hat{e}_2(s);\ \hat{e}_2' = \kappa_o(s) \cdot \hat{e}_1(s); \quad (1)$$

$$\vec{r} = \vec{r}_o(s) + q_1 \cdot \hat{e}_1(s) + q_2 \cdot \hat{e}_2(s);$$

where $f' \equiv \frac{df}{ds}, f'' \equiv \frac{d^2 f}{ds^2} \ldots$ , $K_o(s) = \frac{1}{\rho_o(s)}$ is the curvature of the trajectory, and $\kappa_o(s)$ is its torsion.

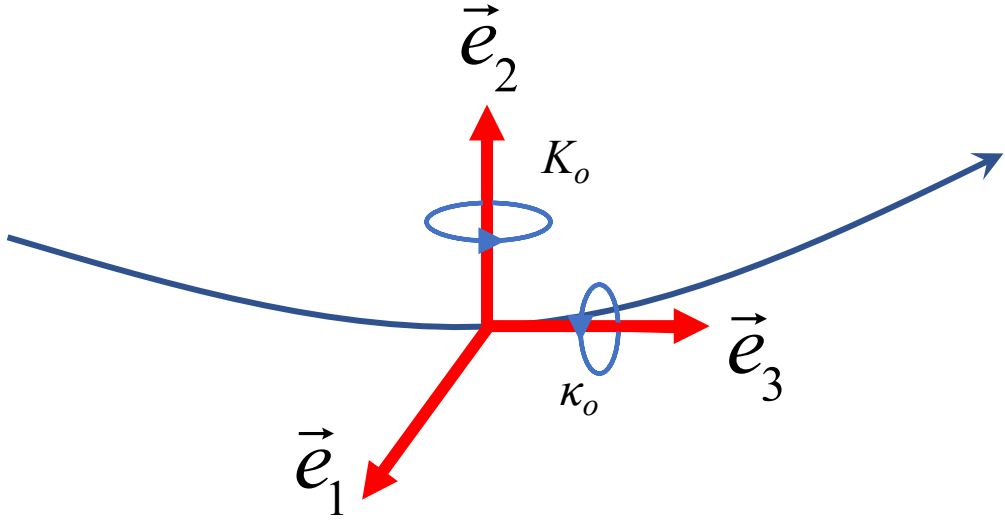

Figure 2. Accelerator coordinate system including curvature $K_o(s)$ and torsion $\kappa_o(s)$.

Any vector $\vec{A}$ can be expanded at azimuth $s$ using this coordinate system as [1]

$$\vec{A} = A_1 \cdot \hat{e}_1 + A_2 \cdot \hat{e}_2 + A_3 \cdot \hat{e}_3;\ A_i = \hat{e}_i \cdot \vec{A}.$$

Further in this paper we will use traditional notations, where $c$ is the speed of the light, $e$, $m$ and $\vec{v}$ are particle's charge, mass and velocity, correspondingly, $\vec{\beta} = \vec{v}/c; \gamma = (1-\vec{\beta}^2)^{-\frac{1}{2}}$ are the relativistic factors, $\vec{p} = \gamma m \vec{v}$ and $\mathbf{E} = \gamma m c^2$ are the mechanical momentum and energy of a particle. and $\varphi, \vec{A}$ are scalar and vector potentials of electromagnetic (EM) field. We will also use subscript "o" to indicate values obtained by variables at the reference trajectory $\vec{r} = \vec{r}_o(s)$ for reference particle with momentum $\vec{p}_o(s) = m \cdot \gamma_o(s) \cdot \vec{v}_o(s)$ and energy $\mathbf{E}_o(s) = \gamma_o(s) \cdot mc^2$ reaching azimuth $s$ at time

$$t_o(s) = \int_0^s \frac{d\zeta}{|\vec{v}_o(s)|}. \tag{2}$$

---

[1] We intentionally do not use contravariant $\vec{A} = A^1 \cdot \hat{e}_1 + A^2 \cdot \hat{e}_2 + A^3 \cdot \hat{e}_3$ and covariant indices $\vec{A} = A_1 \cdot \hat{e}^1 + A_2 \cdot \hat{e}^2 + A_3 \cdot \hat{e}^3$, commonly used in curvilinear coordinate systems, to avoid confusion between squares (cubes) of the value and second (third) contravariant component of a vector. We are also avoiding use of zero's components of 4-vectors, such as $A_o, x_o$ to avoid confusion either with initial condition or values at the reference orbit. This important distinction is not needed and will not be used in the rest of the paper.

In classical Hamiltonian mechanics the time plays role of independent variable and components of particles position, $\vec{r}$, and canonical momenta, $\vec{P}$, are known as canonical pairs, $(q_i, P_i)$. The full set of Canonical pairs and system Hamiltonian fully describe particle's evolution [37-38]

$$H \equiv \mathbf{E} + e\varphi = \sqrt{m^2c^4 + \left(\vec{P} - \frac{e}{c}\vec{A}\right)^2} + e\varphi; \quad \vec{P} = \vec{p} + \frac{e}{c}\vec{A};$$

$$\frac{dq_i}{dt} = \frac{\partial H}{\partial P_i}; \frac{dP_i}{dt} = -\frac{\partial H}{\partial q_i}, \tag{3}$$

Using $s$ as independent variable retains two canonical pairs $(q_1, P_1), (q_2, P_2)$, and generates new third canonical pair: $(-ct, H/c)$. The arrival time of a particle, $t$, and $H$ become $s$-dependent variables with the accelerator Hamiltonian of [34-36]:

$$h_1 = -(1 + K_o q_1)\sqrt{\frac{(H - e\varphi)^2}{c^2} - m^2c^2 - \left(P_1 - \frac{e}{c}A_1\right)^2 - \left(P_2 - \frac{e}{c}A_2\right)^2}$$

$$-\frac{e}{c}(1 + K_o q_1)A_3 + \kappa_o(q_1 P_2 - q_2 P_1). \tag{4}$$

Canonical transformation with the generation function [37]

$$\Phi = q_1 P_1 + q_2 P_2 + q_3 \cdot \frac{H - \mathbf{E}_o(s) - e\varphi(\vec{r}_o(s), t)}{c} + H \cdot (t_o(s) - t)$$

reduces the third canonical pair to $(q_3, P_3)$:

$$q_3 = c(t_o(s) - t); P_3 = \frac{\mathbf{E} - \mathbf{E}_o(s)}{c} + e\frac{\varphi(\vec{r}, t) - \varphi(\vec{r}_o(s), t)}{c}, \tag{5}$$

with zero values for reference particle, $q_3|_o = 0, P_3|_o = 0$, and it reduces the Hamiltonian (4) to

$$h(q, P) = h_1 - \frac{c}{v_o(s)} P_3 + q_3 \frac{d}{ds}\left(\mathbf{E}_o(s) + e\frac{\varphi(\vec{r}_o(s), t)}{c}\right). \tag{6}$$

For compactness, we define a set of notations for coordinates, $q$, and corresponding Canonical momenta, $P$, as well as a phase space vector $\xi$ as

$$q^T = [q_1, q_2, q_3], \quad P^T = [P_1, P_2, P_3], \quad \xi^T = [q^T, P^T], \tag{7}$$

where index $^T$ indicates transposition of matrices, including transferring a column into a row and vice versa. We will call sets $[q_1, q_2, q_3]$ and $[P_1, P_2, P_3]$ phase space coordinates and momenta, correspondingly. Using the definition from Eq. (7), one can write the 3D equations of motion in a compact symplectic form [34] [2]:

$$\frac{d\xi_i}{ds} = S_{ij} \frac{\partial h}{\partial \xi_j} \Leftrightarrow \frac{d\xi}{ds} = \mathbf{S} \frac{\partial h}{\partial \xi};$$

$$\mathbf{S} \equiv [S_{ij}] = \begin{bmatrix} \mathbf{0} & \mathbf{I}_{3\times 3} \\ -\mathbf{I}_{3\times 3} & \mathbf{0} \end{bmatrix}; \mathbf{I}_{3\times 3} = \begin{bmatrix} 1 & 0 & 0 \\ 0 & 1 & 0 \\ 0 & 0 & 1 \end{bmatrix}; \quad \mathbf{S}^2 = -\mathbf{I}_{6\times 6}; \tag{8}$$

where $i, j = (1, 2, 3)$, $\mathbf{0}$ is a 3x3 zero matrix (see Appendix A for further discussion). The number of components can be proportionally reduced for the 2D and 1D cases.

Motion of particles is determined by the initial conditions[3]

$$\underline{q} \equiv q(s=0), \quad \underline{P} \equiv P(s=0), \quad \underline{\xi} \equiv \xi(s=0),$$

with solved equations of motion

$$q = q(\underline{q}, \underline{P}, s), \quad P = P(\underline{q}, \underline{P}, s), \quad \xi = \xi(\underline{\xi}, s), \tag{9}$$

representing the Canonical transformation from $(\underline{q}, \underline{P})$ to $(q, P)$ [37]. The inverse transformation

$$\underline{q} = \underline{q}(q, P, s), \quad \underline{P} = \underline{P}(q, P, s), \quad \underline{\xi} = \underline{\xi}(\xi, s), \tag{10}$$

not only exists but also is a Canonical transformation from $(q, P)$ to $(\underline{q}, \underline{P})$. The transformation (10) results in zero Hamiltonian for the set of canonical variables $(\underline{q}, \underline{P})$:

$$\underline{h}(\underline{q}, \underline{P}, s) = 0, \text{ [4]}$$

which is a traditional way of solving evolution for the background distribution function upon which instability can develop. This method is called "the variation of initial values" in analytical mechanics [38] or "the method of trajectories" in plasma physics [39]. By assuming that the solutions for the self-consistent trajectories in Eqs. (8-10) are known, it allows us to remove the dynamic terms and reduce the Vlasov equations to the ones comprising of the perturbation terms only. Hence, we assume that the solution for an unperturbed distribution function $f_o$ is known and satisfies the self-consistent Vlasov equation [5] [40]:

$$\frac{\partial f_o(\xi,s)}{\partial s} + S_{ik}\frac{\partial f_o(\xi,s)}{\partial \xi_i}\frac{\partial h_o(\xi,s)}{\partial \xi_k} = 0;$$

$$\underline{f_o}(\underline{\xi}) \equiv f_o(\xi,s=0) \Rightarrow f_o(\xi,s) = \underline{f_o}(\underline{\xi}(\xi,s)).$$

(11)

---

[2] Further in the paper we will use Einstein's convention of summation by repeated indices, e.g.,
$a_i b_i \equiv \sum_i a_i b_i; \quad a_i b_{ki} c_{nk} \equiv \sum_i \sum_k a_i b_{ki} c_{nk}.$

[3] We will continue using underscore $\underline{f}$ for initial values at $s=0$: $\underline{f} \equiv f(...., s=0)$.

[4] This transformation can leave $\underline{h}(\underline{q}, \underline{P}, s) = f(s)$, which can be easily removed by a Canonical transformation $F = \underline{q}_i \underline{P}_i - \int_0^s f(z) dz$.

Let's now consider an infinitesimally small perturbation of the distribution function, $\tilde{f}$,

$$f(\xi,s) = f_o(\xi,s) + \tilde{f}(\xi,s); \ |\tilde{f}(\xi,s)| \ll |f_o(\xi,s)|,$$

(12)

e.g., $|\tilde{f}(\xi,s)| = O(\varepsilon)|f_o(\xi,s)|; \ \varepsilon \ll 1$, and the corresponding weak perturbation in the Hamiltonian:

$$h(\xi,s) = h_o(\xi,s) + \tilde{h}(\xi,s); \ \tilde{h}(\xi,s) = O(\varepsilon).$$

(13)

Applying Canonical transformation (10) we reduce the Hamiltonian (13) to the perturbation term

$$\underline{h}(\underline{\xi},s) = \underline{\tilde{h}}(\underline{\xi},s) \equiv \tilde{h}(\xi(\underline{\xi},s),s),$$

(14)

With the Vlasov equations for the corresponding variation of the initial distribution function $\underline{\tilde{f}}$:

$$\underline{f}(\underline{\xi},s) = \underline{f_o}(\underline{\xi}) + \underline{\tilde{f}}(\underline{\xi},s); \ \tilde{f}(\xi,s) \equiv \underline{\tilde{f}}(\underline{\xi}(\xi,s),s);$$

$$\frac{\partial \underline{\tilde{f}}}{\partial s} + S_{ik}\frac{\partial \underline{f_o}}{\partial \underline{\xi}_j}\frac{\partial \underline{\tilde{h}}}{\partial \underline{\xi}_k} + S_{ik}\frac{\partial \underline{\tilde{f}}}{\partial \underline{\xi}_j}\frac{\partial \underline{\tilde{h}}}{\partial \underline{\xi}_k} = 0.$$

(15)

Next standard step is the linearization of the Vlasov equation by recognizing that the third term in second line in Eq. (15) is on the order of $O(\varepsilon^2)$:

$$\frac{\partial \underline{\tilde{f}}}{\partial s} + S_{ik}\frac{\partial \underline{f_o}}{\partial \underline{\xi}_i}\frac{\partial \underline{\tilde{h}}}{\partial \underline{\xi}_k} = -S_{ik}\frac{\partial \underline{\tilde{f}}}{\partial \underline{\xi}_i}\frac{\partial \underline{\tilde{h}}}{\partial \underline{\xi}_k} = O(\varepsilon^2) \to 0;$$

$$\frac{\partial \underline{\tilde{f}}}{\partial s} + \frac{\partial \underline{f_o}}{\partial \underline{q}_j}\frac{\partial \underline{\tilde{h}}}{\partial \underline{P}_j} - \frac{\partial \underline{f_o}}{\partial \underline{P}_j}\frac{\partial \underline{\tilde{h}}}{\partial \underline{q}_j} = 0.$$

(16)

---

[5] Self-consistent distribution function, which we use as the known background, would include all macroscopic collective effects such as SC and wake-fields induced by the bunch. The self-consistent Hamiltonian would have functional dependence on the initial beam distribution $\underline{f_o}(\underline{\xi})$, e.g., $h = h_o(q, P, s, \underline{f_o}(\underline{\xi}))$. This fact does not change the validity and applicability of the Vlasov equation (11).

It is known that a generic 3D evolution of a finite size charged beam is analytically intractable. Rare exceptions, such as non-physical but self-consistent Kapchinsky-Vladimirsky (KV) distribution [41], only attest to the case. Several further assumptions are needed to analytically derive solvable equation[6].

One typical simplification used in the theory of beam instabilities is an assumption of homogenous background density. While this approach is not applicable for all collective effects in a beam with finite sizes, it has limited applicability for analyzing evolution of perturbations with periods significantly smaller than the typical scales of the beam's uniformity.

It is intuitively understandable that scales of the beam uniformity $a_i$

$$\left|\frac{\partial f_o}{\partial q_i}\right| \propto \frac{f_o}{a_i} \tag{17}$$

define the scale of the perturbations when the homogenous background density can be used as a good approximation. Appendix B has detailed studies of these requirements. It can be summarized as follows: the k-vector, $k^T = [k_1, k_2, k_3]$ of Fourier component with exponential factor $e^{ik^T q} = e^{i\vec{k} \cdot \vec{q}}$ must satisfy following conditions [7]:

$$a_{1,2} \cdot \sqrt{(\beta_o \gamma_o)^2 (k_1^2 + k_2^2) + k_3^2} \gg \beta_o \gamma_o; \quad a_3 \cdot \sqrt{(\beta_o \gamma_o)^2 (k_1^2 + k_2^2) + k_3^2} \gg 1. \tag{18}$$

Since we are considering a generic accelerator, which can include beam's focusing and bending of its trajectory, acceleration, compression, or decompression, we shall also assume that changes in the beam and the accelerator parameters at the scale of the density modulation are negligible:

$$\left|\vec{\nabla} g\right| \ll \left|\vec{k}\right| |g|; \tag{19}$$

where $g$ is any generic parameter of the accelerator, including but not limited to the beam's energy, velocity, sizes, the accelerator EM fields, the curvature and the torsion of the reference beam trajectory.

Nonlinearity of the transfer map $\xi = \mathbf{M} : X$ could cause distortions resulting in coupling between Fourier harmonics of the density perturbation. As shown in Appendix B, such coupling would make further analytical evaluation impossible. Hence, we are considering linearization of the self-consistent symplectic map $\xi = \mathrm{M}(s):\underline{\xi}$ with small deviations $\Delta\xi(s)$ nearby a selected phase-space trajectory $\underline{\xi}_o(s) = \mathrm{M}(s):\underline{\xi}_o$ [8]:

$$\underline{\xi} = \underline{\xi}_o + \Delta\underline{\xi}, \quad \mathbf{M}_{\underline{\xi}_o}(s) = \left.\frac{\partial \xi}{\partial \underline{\xi}}\right|_{\underline{\xi}=\underline{\xi}_o} \equiv \left.\frac{\partial}{\partial \underline{\xi}}\left(\mathrm{M}(s):\underline{\xi}\right)\right|_{\underline{\xi}=\underline{\xi}_o};$$

$$\xi(s) = \xi_o(s) + \Delta\xi(s) = \mathrm{M}(s):\underline{\xi}_o + \mathbf{M}_{\underline{\xi}_o}(s)\Delta\underline{\xi} + O\left(\left|\Delta\underline{\xi}\right|^2\right),$$

with linear map (matrix) implicitly depending on the starting point of phase-space trajectory, $\underline{\xi}_o$. Rewriting this map expansion as

$$\begin{bmatrix} q(s) \\ P(s) \end{bmatrix} = \begin{bmatrix} \mathrm{M}_q(s):\underline{\xi}_o \\ \mathrm{M}_P(s):\underline{\xi}_o \end{bmatrix} + \begin{bmatrix} \mathbf{A} & \mathbf{B} \\ \mathbf{C} & \mathbf{D} \end{bmatrix} \begin{bmatrix} \Delta\underline{q} \\ \Delta\underline{P} \end{bmatrix} + O\left(\left|\Delta\underline{q}\right|^2, \left|\Delta\underline{P}\right|^2\right),$$

we can define area of the phase space, $\Omega$, where nonlinear distortion and coupling between Fourier harmonics can be neglected

$$\delta q = \mathrm{M}_q(s):\left(\underline{\xi}_o + \Delta\underline{\xi}\right) - \mathrm{M}_q(s):\underline{\xi}_o - \mathbf{A}\Delta\underline{q} - \mathbf{B}\Delta\underline{P};$$
$$\left|\vec{k}\cdot\delta\vec{q}\right| \ll 1; \quad \{\Delta\underline{q}, \Delta\underline{P}\} \in \Omega, \tag{20}$$

---

[6] Typically, the combination of Vlasov and Maxwell equations is not directly solvable because it contains partial derivatives.

[7] Where it is convenient, we will use objects such as $\vec{x} = \sum_{i=1}^{3} \hat{e}_i x_i$ for $x^T = [x_1, x_2, x_3]$, with product defined as $\vec{x} \cdot \vec{y} = \sum_{i=1}^{3} x_i y_i \equiv x^T \cdot y \equiv y^T \cdot x$. It is important to note that these vectors are not real 3D vectors. We will also use compact notation for convolution of objects and matrices: $\vec{\mathbf{A}} \cdot \vec{y} = \sum_{i=1}^{3} \hat{e}_i \sum_{j=1}^{3} A_{ij} y_j; \quad \vec{x} \cdot \vec{\mathbf{A}} \cdot \vec{y} = \sum_{i=1}^{3} \sum_{j=1}^{3} A_{ij} x_i y_j$.

The transfer map should be evaluated self-consistently, including macroscopic collective effects. Further in the paper we will drop $\Delta$ in front of $\xi$ and will use *6x6* symplectic transport matrix [34-36]

$$\xi = \mathbf{M}(s)\underline{\xi}; \quad \underline{\xi} = \mathbf{M}^{-1}(s)\xi; \quad \mathbf{M}(0) = \mathbf{I}_{6\times 6};$$
$$\mathbf{M}^T\mathbf{S}\mathbf{M} = \mathbf{M}\mathbf{S}\mathbf{M}^T = \mathbf{S} \Rightarrow \det\mathbf{M} = 1; \quad \mathbf{M}^{-1} = -\mathbf{S}\mathbf{M}^T\mathbf{S}, \tag{21}$$

in the vicinity $\Omega$ of initial condition $\underline{\xi}_o$. It is convenient to identify four *3x3* block-matrices in the transport and inverse matrices:

$$\begin{bmatrix} q \\ P \end{bmatrix} = \mathbf{M}(s)\begin{bmatrix} \underline{q} \\ \underline{P} \end{bmatrix}; \quad \begin{bmatrix} \underline{q} \\ \underline{P} \end{bmatrix} = \mathbf{M}^{-1}(s)\begin{bmatrix} q \\ P \end{bmatrix};$$
$$\mathbf{M} = \begin{bmatrix} \mathbf{A} & \mathbf{B} \\ \mathbf{C} & \mathbf{D} \end{bmatrix}; \quad \mathbf{M}^{-1} = -\mathbf{S}\mathbf{M}^T\mathbf{S} = \begin{bmatrix} \mathbf{D}^T & -\mathbf{B}^T \\ -\mathbf{C}^T & \mathbf{A}^T \end{bmatrix}; \tag{22}$$

providing explicit connections between the local and initial coordinates and momenta:

---

[8] Nonlinearity of the map would result in a nonlinear, position-deptendent transformation of the *k*-vector canceling advantages offered by the Fourier transformation. Generally speaking, a nonlinear map would not allow us to extend our studies beyond Eq.(17).

$$\underline{q} = \mathbf{A}q + \mathbf{B}\underline{P}; \quad P = \mathbf{C}q + \mathbf{D}\underline{P};$$
$$q = \mathbf{D}^T\underline{q} - \mathbf{B}^T P; \quad \underline{P} = -\mathbf{C}^T\underline{q} + \mathbf{A}^T P. \tag{23}$$

It's worth noticing that in this notation [34] three degrees of motion are decoupled when all four 3x3 matrices, **A, B, C** and **D** are diagonal (see Appendix A for more details).

Matrix **A** plays a special role for this instability since its determinant represents the degree of the three-dimensional bunch compression:

$$\rho(q,s) = e\int_{-\infty}^{\infty} dP^3 f(q,P) = \frac{e}{\det \mathbf{A}}\int_{-\infty}^{\infty} d\underline{P}^3 \underline{f}\left(\mathbf{A}^{-1}(q - \mathbf{B}\underline{P}), \underline{P}\right). \tag{24}$$

where we used one of Eq. (23) to connect local beam densities (at azimuth *s*) with their initial values at *s=0*:

$$\underline{P} = -\mathbf{C}^T\underline{q} + \mathbf{A}^T P \Rightarrow P = \left(\mathbf{A}^T\right)^{-1}\left(\underline{P} + \mathbf{C}^T\underline{q}\right) \Rightarrow dP^3\big|_{\underline{q}=const} = \frac{1}{\det \mathbf{A}} d\underline{P}^3\big|_{q=const};$$
$$q = \mathbf{A}\underline{q} + \mathbf{B}\underline{P} = const \Rightarrow \underline{q} = \mathbf{A}^{-1}(q - \mathbf{B}\underline{P}); \tag{25}$$
$$f\left(q, \left(\mathbf{A}^T\right)^{-1}\left(\underline{P} + \mathbf{C}^T\underline{q}\right)\right) = \underline{f}\left(\mathbf{A}^{-1}(q - \mathbf{B}\underline{P}), \underline{P}\right).$$

One of the important consequences of using the assumption of a homogeneous background density, described by background distribution function of $\underline{f}_o(\underline{P})$, results in requirement of $\det \mathbf{A} > 0$. Because of the assumption of a homogeneous background density, beam is effectively infinite, and its density would become infinitely large, e.g. unphysical, when $\det \mathbf{A} = 0$. While it is already indicated by $\det \mathbf{A}$ in the denominator in Eq.(24) for the particle density $\rho(q,s) = \frac{1}{\det \mathbf{A}}\int_{-\infty}^{\infty} d\underline{P}^3 \underline{f}_o(\underline{P})$, this is most evident in a 1D case

$$\mathbf{M} = \begin{bmatrix} m_{11} & m_{12} \\ m_{21} & m_{22} \end{bmatrix}; \quad \mathbf{M}^{-1} = -\mathbf{S}\mathbf{M}^T\mathbf{S} = \begin{bmatrix} m_{22} & -m_{12} \\ -m_{21} & m_{11} \end{bmatrix},$$

where $m_{11}$ plays a role of the det **A** and the change in the line density can be easily expressed as

$$\rho(q,s) = \int_{-\infty}^{\infty} \underline{f}_o(-m_{21}q + m_{11}\underline{P}) d\underline{P} = \frac{1}{m_{11}}\int_{-\infty}^{\infty} \underline{f}_o(\underline{P}) d\underline{P} = \frac{n_o}{m_{11}}. \tag{26}$$

We discuss consequences and solution for handling cases of $\det \mathbf{A} \to 0$ in **Section V** of this paper.

As shown in Appendices C and D, density perturbation will generate additional potentials of the EM field resulting in the following perturbation of the accelerator Hamiltonian (see equations (E5)):

$$\tilde{h} = \frac{4\pi e^2}{c}\int \frac{\tilde{\rho}_{\vec{k}} e^{i\vec{k}\vec{q}} dk^3}{\gamma_o^2\beta_o^2 \vec{k}_\perp^2 + k_3^2}; \quad \delta\frac{d\vec{P}}{ds} = -\frac{\partial \tilde{h}}{\partial \vec{q}} = -\frac{4\pi e^2}{c}\int \frac{i\vec{k} \cdot \tilde{\rho}_{\vec{k}} e^{i\vec{k}\vec{q}} dk^3}{\gamma_o^2\beta_o^2 \vec{k}_\perp^2 + k_3^2}, \tag{27}$$

where $\vec{k}\cdot\vec{q} = k^T q = \sum_{i=1}^{3} k_i q_i$; $dk^3 \equiv \prod_{i=1}^{3} dk_i$; $\vec{k}_\perp^2 = k_1^2 + k_2^2$. We can easily connect $\tilde{\rho}_{\vec{k}}$ at a location $s$ with the Fourier harmonic of $\tilde{f}$ and $\tilde{\underline{f}}$. Considering conservation of the phases-space volume $dq^3 dP^3 = \det \mathbf{M} \cdot d\underline{q}^3 d\underline{P}^3 = d\underline{q}^3 d\underline{P}^3$ and conservation of the phase space density $\tilde{f}(X,s) \equiv \tilde{\underline{f}}(\xi(\underline{\xi},s),s)$ we get:

$$\tilde{\rho}_{\vec{k}} \equiv \tilde{\rho}(s,\vec{k}) = \frac{1}{(2\pi)^3} \iint \tilde{f}(q,P,s) e^{-i\vec{k}\cdot\vec{q}} \, dq^3 \, dP^3 = $$
$$\frac{1}{(2\pi)^3} \iint \tilde{\underline{f}}(\underline{q},\underline{P},s) e^{-i\vec{k}\cdot\vec{q}(\underline{\xi},t)} \, d\underline{q}^3 \, d\underline{P}^3 = \frac{1}{(2\pi)^3} \iint \tilde{\underline{f}}(\underline{q},\underline{P},s) \cdot e^{-i\vec{k}\cdot(\vec{\mathbf{A}}\underline{q}+\vec{\mathbf{B}}\cdot\underline{P})} \, d\underline{q}^3 \, d\underline{P}^3, \quad (28)$$

where we used $\vec{q} = \vec{\mathbf{A}} \cdot \underline{\vec{q}} + \vec{\mathbf{B}} \cdot \vec{\underline{P}}$ as a object equivalent of $q = \mathbf{A}\underline{q} + \mathbf{B}\underline{P}$ in Eq. (23).

It is natural place to discuss evolution of wavenumbers. As can be seen from Eq(28), that $\vec{k}$-vector of density modulation at azimuth $s$ is connected to that at $s=0$:

$$\underline{k} = k(s=0); \quad \underline{k}^T = k^T(s) \cdot \mathbf{A}(s) \Rightarrow k^T(s) = \underline{k}^T \cdot \mathbf{A}^{-1}(s). \quad (29)$$

or in vector form [9]:

$$\vec{k}(s) = \underline{\vec{k}} \cdot \vec{\mathbf{A}}^{-1}(s); \quad \vec{k}(s) \cdot \vec{q} = \underline{\vec{k}} \cdot \left(\underline{\vec{q}} + \overrightarrow{\mathbf{A}^{-1}\mathbf{B}} \cdot \underline{\vec{P}}\right).$$

It means that matrix $\mathbf{A}$, the spatial components of the transport matrix, also defines evolution of the $k$-vector with initial value of $\underline{\vec{k}} = \vec{k}(s=0)$:

$$k^T(s) = [k_1(s), k_2(s), k_3(s)]; \quad \underline{k} = \mathbf{A}^T k(s) \Leftrightarrow k(s) = (\mathbf{A}^T)^{-1} \underline{k}. \quad (30)$$

We can assume, without a loss of generality, that the initial back-ground distribution is an arbitrary integrable function of momenta [10]:

$$\underline{f}_o \Rightarrow n_o f_o(\underline{P}); \int_{-\infty}^{\infty} f_o(\underline{P}) d\underline{P}^3 = 1; \quad n_o = \frac{j_o}{ec}, \quad (31)$$

where $j_o$ is the initial beam current density. It is important to note that in contrast with velocity-dependent spatial density of the beam, $n_l = j_o/ev_o$, the $n_o = \beta_o n_l$ has a well-defined finite value.

---

[9] For compactness, in places where it cannot cause confusion, we omit the explicit indication of $s$-dependence, for example using $\overrightarrow{\mathbf{A}^{-1}\mathbf{B}}$ instead of $\overrightarrow{\mathbf{A}(s)^{-1}\mathbf{B}(s)}$.

[10] For plasma to remain uniform the distribution must have form of $f(\underline{P}+\mathbf{M}\underline{q})$. Initial linear correlations between $\underline{P}$ and $\underline{q}$ can be incorporated into the transport matrix (22).

Using relations in Eq.(25) and taking into account that $\delta\left(\frac{dq_j}{ds}\right) = \frac{\partial \tilde{h}(\xi,s)}{\partial P_j} = 0$, we can rewrite Vlasov equation (16) as follows:

$$\underline{P} = \mathbf{A}^T P - \mathbf{C}^T q; \, d\underline{P}_i = A_{ji} dP_j - C_{ji} dq_j;$$

$$\frac{\partial \tilde{f}}{\partial s} = -n_o \frac{\partial f_o}{\partial \underline{P}_i} \frac{\partial \underline{P}_i}{\partial P_j}\delta\left(\frac{dP_j}{ds}\right) - n_o \frac{\partial f_o}{\partial \underline{P}_i}\frac{\partial \underline{P}_i}{\partial q_j}\delta\left(\frac{dq_j}{ds}\right) = n_o \frac{\partial f_o}{\partial \underline{P}_i} A_{ji} \frac{\partial \tilde{h}}{\partial q_j},$$

and introducing the perturbation Hamiltonian (27) to arrive to the self-consistent Vlasov equations:

$$\frac{\partial \tilde{f}}{\partial s} = n_o \frac{\partial f_o}{\partial \underline{P}_i} A_{ji}(s) \mathbf{F}_j(q,s); \, \mathbf{F}_j(q,s) = \frac{\partial \tilde{h}}{\partial q_j} = \frac{4\pi e^2}{c}\int \frac{ik_j \cdot \tilde{\rho}_{\vec{k}} e^{i\vec{k}\cdot\vec{q}} dk^3}{\gamma_o^2 \beta_o^2 \vec{k}_\perp^2 + k_3^2}. \tag{32}$$

Applying Fourier transform $\underline{f}_{\vec{k}}(\underline{P},s) = \frac{1}{(2\pi)^3}\int_{-\infty}^{\infty} \underline{f}(q,\underline{P},s)e^{-i\vec{k}\cdot\vec{q}}dq^3$ to this equation we get:

$$\frac{\partial \tilde{f}_{\vec{k}}}{\partial s} = \frac{n_o}{(2\pi)^3}\frac{\partial f_o}{\partial \underline{P}_i} A_{ji}(s)\int dq^3 e^{-i\vec{k}\cdot\vec{q}} \mathbf{F}_j(q,s). \tag{33}$$

The latter must be evaluated at $\vec{P} = const$ using the established relations between the $k$-vectors (30):

$$\mathbf{F}_{\vec{k}} = \frac{1}{(2\pi)^3}\int e^{-i\vec{k}\cdot\vec{q}} \mathbf{F}(q,s)dq^3 \bigg|_{\underline{P}=const} = \frac{4\pi e^2}{c}\int \frac{i\vec{k}\tilde{\rho}_{\vec{k}} dk^3}{\gamma_o^2 \beta_o^2 \vec{k}_\perp^2 + k_3^2}\cdot\frac{1}{(2\pi)^3}\int e^{i\vec{k}\cdot\vec{q}}e^{-i\underline{\vec{k}}\cdot\vec{q}} dq^3;$$

$$\frac{1}{(2\pi)^3}\int e^{i\vec{k}\vec{q}}e^{-i\underline{\vec{k}}\cdot\vec{q}} dq^3 = \frac{e^{i\vec{k}\cdot\overset{\leftrightarrow}{\mathbf{B}}\cdot\vec{P}}}{(2\pi)^3}\int e^{i(\vec{k}\cdot\mathbf{A}-\underline{\vec{k}})\cdot\vec{q}} dq^3 = e^{i\vec{k}\cdot\overset{\leftrightarrow}{\mathbf{B}}\cdot\vec{P}}\delta(\vec{k}\cdot\mathbf{A} - \underline{\vec{k}}) = \frac{e^{i\underline{\vec{k}}\cdot\overline{\mathbf{A}^{-1}\mathbf{B}}\cdot\vec{P}}}{\det\mathbf{A}}\delta(\vec{k} - \underline{\vec{k}}\cdot\mathbf{A}^{-1}),$$

resulting in

$$\frac{\partial \tilde{f}_{\vec{k}}(\underline{P},s)}{\partial s} = \frac{4\pi n_o e^2}{c}\cdot\frac{\tilde{\rho}(s,\vec{k}(s))}{\gamma_o(s)^2 \beta_o(s)^2 \vec{k}_\perp(s)^2 + k_3(s)^2}\frac{e^{i\underline{\vec{k}}\cdot\overline{\mathbf{A}^{-1}(s)\mathbf{B}(s)}\cdot\vec{P}}}{\det\mathbf{A}(s)}\left(i\underline{k}_i \frac{\partial f_o}{\partial \underline{P}_i}\right), \tag{34}$$

where we took into account that $k_j(s) A_{ji}(s) = \underline{k}_i$.

This equation can be easily integrated:

$$\tilde{f}_{\vec{k}}(\underline{P},s) = \tilde{f}_{\vec{k}}(\underline{P},0) + \frac{4\pi i n_o e^2}{c}\left(\underline{k}_i \frac{\partial f_o}{\partial \underline{P}_i}\right)\int_o^s \frac{e^{i\underline{\vec{k}}\cdot\overline{\mathbf{A}^{-1}(\zeta)\mathbf{B}(\zeta)}\cdot\vec{P}}}{\det\mathbf{A}(\zeta)}\frac{\tilde{\rho}_{\vec{k}}(\zeta)d\zeta}{\gamma_o(\zeta)^2 \beta_o(\zeta)^2 \vec{k}_\perp^2(\zeta) + k_3^2(\zeta)}, \tag{35}$$

where $\tilde{f}_{\vec{k}}(\underline{P},0) \equiv \tilde{f}_{\vec{k}(0)}(\underline{P}, s=0)$ is a Fourier harmonic of the initial perturbation. Rewriting (28) as

$$\tilde{\rho}\left(s,\vec{k}(s)\right) = \int \underline{\tilde{f}_{\vec{k}}}\left(\underline{P},s\right) \cdot e^{-i\underline{\vec{k}} \cdot \overline{\mathbf{A}(s)^{-1}\mathbf{B}(s)} \cdot \underline{\vec{P}}} \, d\underline{P}^3, \tag{36}$$

turns Eq. (35) into a directly solvable integral equation:

$$\tilde{\rho}\left(s,\vec{k}(s)\right) = \tilde{\rho}_{o\underline{k}}(s) + \frac{4\pi i e^2 n_o}{c} \int_o^s \frac{\tilde{\rho}\left(\zeta,\vec{k}(\zeta)\right)d\zeta}{\det \mathbf{A}(\zeta)} \int \frac{e^{i\left(\vec{k}(\zeta)\cdot\vec{\mathbf{B}}(\zeta) - \vec{k}(s)\cdot\vec{\mathbf{B}}(s)\right)\cdot\underline{\vec{P}}}}{\gamma_o(\zeta)^2 \beta_o(\zeta)^2 \vec{k}_\perp^2(\zeta) + k_3^2(\zeta)} \underline{k}_i \frac{\partial \underline{f_o}}{\partial \underline{P}_i} d\underline{P}^3;$$
$$\tilde{\rho}_{o\underline{k}}(s) = \int e^{-i\underline{\vec{k}}(s)\cdot\vec{\mathbf{B}}(s)\cdot\underline{\vec{P}}} \underline{\tilde{f}_{\vec{k}}}\left(\underline{P},0\right) d\underline{P}^3. \tag{37}$$

While this equation already can be used for evaluation of the instability, it can be further simplified by eliminating convolution $\sum_{i=1}^{3} \underline{k}_i \frac{\partial \underline{f_o}}{\partial \underline{P}_i}$. Integrating by parts

$$\int \frac{\partial \underline{f_o}}{\partial \underline{P}_i} \phi \, d\underline{P}_i = \underline{f_o} \phi \Big|_{\underline{P}_i=-\infty}^{\underline{P}_i=\infty} - \int \underline{f_o} \frac{\partial \phi}{\partial \underline{P}} d\underline{P} \tag{38}$$

and $\underline{f_o}\left(\underline{P}_i = \pm\infty\right) = 0$ we get:

$$\sum_{i=1}^{3} \underline{k}_i \frac{\partial}{\partial \underline{P}_i} e^{i\underline{\vec{k}}\cdot\left(\vec{\mathbf{U}}(\zeta)-\vec{\mathbf{U}}(s)\right)\cdot\underline{\vec{P}}} = -i\left(u(s)-u(\zeta)\right);$$
$$u(\zeta) = \vec{k} \cdot \vec{\mathbf{B}}(\zeta) \cdot \underline{\vec{k}} \equiv \sum_{i,j} \mathbf{B}_{ij}(\zeta) \cdot k_i \cdot \underline{k}_j = \sum_{i,j} \left[\mathbf{A}(\zeta)^{-1}\mathbf{B}(\zeta)\right]_{ij} \cdot \underline{k}_i \cdot \underline{k}_j \tag{39}$$

Combining Eqs. (37) and (38) brings us to the final form of the integral equation for 3D SC instability:

$$\tilde{\rho}\left(s,\vec{k}(s)\right) = -\int_o^s \tilde{\rho}\left(\zeta,\vec{k}(\zeta)\right) \cdot K(\zeta)\left(u(s)-u(\zeta)\right) L_d(s,\zeta) d\zeta + \tilde{\rho}_{o\underline{k}}(s);$$
$$K(\zeta) = \frac{4\pi n_o e^2}{c \det \mathbf{A}(\zeta) \upsilon(\zeta)}; \quad L_d\left(\vec{k},s,\zeta\right) = \int e^{i\left(\vec{k}(\zeta)\cdot\vec{\mathbf{B}}(\zeta) - \vec{k}(s)\cdot\vec{\mathbf{B}}(s)\right)\cdot\underline{\vec{P}}} \underline{f_o}(\underline{P}) d\underline{P}^3; \tag{40}$$
$$u(\zeta) = \vec{k}(\zeta) \cdot \vec{\mathbf{B}}(\zeta) \cdot \underline{\vec{k}} \equiv \vec{k}(\zeta) \cdot \vec{\mathbf{U}}(\zeta) \cdot \underline{\vec{k}}; \quad \upsilon(s) = \gamma_o(s)^2 \beta_o(s)^2 \vec{k}_\perp(s)^2 + k_3(s)^2;$$

which can be solved numerically for any accelerator. Here we defined $\mathbf{U} = \mathbf{A}^{-1}\mathbf{B}$.

It is important to note that in the kernel of the integral equation (40) there is only one term, the Landau damping, $L_d$, which depends on both the value and direction of the k-vector. The $u/\upsilon$ is defined by the geometry (e.g. direction of the initial k-vector) and the components of the accelerator transport matrix in a form of matrix $\mathbf{U} = \mathbf{A}^{-1}\mathbf{B}$ and s-dependent denominator:

$$\frac{u(s)-u(\zeta)}{\upsilon(\zeta)} = \frac{\vec{\vartheta} \cdot \left(\vec{\mathbf{U}}(\zeta)-\vec{\mathbf{U}}(s)\right) \cdot \vec{\vartheta}}{\gamma_o(s)^2 \beta_o(s)^2 \vec{\vartheta}_\perp(s)^2 + \vec{\vartheta}_3(s)^2}; \vec{\vartheta}(s) = \frac{\vec{k}(s)}{|\vec{k}|}; \underline{\vec{\vartheta}} = \vec{\vartheta}(0) = \frac{\underline{\vec{k}}}{|\vec{k}|};$$

We show in Eq.(A3) of Appendix A that $\mathbf{AB}^T = \mathbf{BA}^T$, which also means that $\mathbf{U} = \mathbf{A}^{-1}\mathbf{B}$ is a symmetric matrix.

The most non-trivial construction is actually $\upsilon(\zeta)$, which is the result of the asymmetry between the longitudinal and transverse degrees of freedom introduced by Lorentz transformation:

$$\upsilon(s) = \vec{k} \cdot \vec{\mathbf{G}}(s) \cdot \vec{k}; \ddot{\mathbf{G}} = \mathbf{A}^{-1} \begin{bmatrix} \gamma_o^2 \beta_o^2 & 0 & 0 \\ 0 & \gamma_o^2 \beta_o^2 & 0 \\ 0 & 0 & 1 \end{bmatrix} (\mathbf{A}^{-1})^T. \tag{41}$$

Furthermore, the convolution $u(s) = \underline{\vec{k}} \cdot \ddot{\mathbf{U}}(\zeta) \cdot \underline{\vec{k}}$ has important non-trivial properties that it is a nonnegative monotonic function of $s$ with positive derivative (see Eq.(A11) in Appendix A):

$$u(s) \geq 0; \; u'(s) > 0. \tag{42}$$

Generally speaking, for a beam with an arbitrary momentum spread Eq. (40) cannot be either evaluated analytically or reduced in complexity. But physical nature of various terms can be identified by considering specific cases. For example, the integral over the momenta, known as Landau damping, can be easily evaluated for Gaussian distribution:

$$\underline{f}_o(P) = \prod_{i=1}^{3} \frac{1}{\sqrt{2\pi}\sigma_i} \exp\left(-\frac{P_i^2}{2\sigma_i^2}\right) \tag{43}$$

generating exponential term

$$L_d = \int e^{i(\vec{\eta}(\zeta) - \vec{\eta}(s)) \cdot \vec{P}} F_o(P) dP^3 = \prod_{i=1}^{3} \exp\left(-\frac{\sigma_i^2 (\eta_i(\zeta) - \eta_i(s))^2}{2}\right); \; \vec{\eta}(\zeta) = \vec{k}(\zeta) \cdot \ddot{\mathbf{B}}(\zeta) \tag{44}$$

corresponding to the decay of the modulation during the interval $(\xi, s)$.

To conclude this section, we would like to summarize that equation (40) is the most general equation that describes evolution of the high-frequency modulation in beams driven by space charge effects. It can describe all space-charge driven instabilities from one-dimensional to three-dimensional. For example, it is easy to show that longitudinal microwave instability can be also described by this equation under number of simplified assumptions. Specifically, conventional theory of longitudinal microwave instability assumes that in straight sections the longitudinal motion is frozen and energy modulation resulted from the accumulated SC forces is transferred into density by $R_{56}$ of a magnetic system. Furthermore, SC is frequently neglected in the bending magnetic system. Hence, Eq. (40) is a universal equation for description of instabilities driven by the SC.

### III. Reduction to an ordinary differential equation

In this section, we review some specific cases when we can reduce linear integral equation (40) to a second-order ordinary differential equation (ODE).

Let's consider cases when the Landau damping term allows separation of variables $s$ and $\zeta$[11]:

$$L_d(s,\zeta) = \Lambda(\zeta)\Lambda^{-1}(s); \Lambda(s) = e^{-\phi(s)}; \quad [12] \quad (45)$$

and the integral equation (40) becomes:

$$\tilde{q}(s) = -\int_o^s \tilde{q}(\zeta) K(\zeta)(u(s) - u(\zeta))d\zeta + \tilde{q}_o(s); \quad (46)$$

for a scaled density modulation $\tilde{q}(s) = e^{\phi(s)} \tilde{\rho}(s, \vec{k}(s))$. Combination of first and second derivatives of Eq.(46) transfers it into the second order ODE:

$$\tilde{q}'' - \alpha' \cdot \tilde{q}' + Ku' \cdot \tilde{q} = \tilde{q}_o'' - \alpha' \cdot \tilde{q}_o';$$
$$\tilde{q}_o(s) = e^{\phi(s)}\tilde{\rho}_{\vec{k}0}(s); \; \alpha = \ln\frac{u'}{u_o'}; \; u_o' = u'(0), \quad (47)$$

where we used fact that $u'>0$. This equation can be also reduced to inhomogeneous Hill's equation

$$\hat{q}'' + \hat{K}(s)\hat{q} = \varsigma(s); \; \hat{K}(s) = K(s)u'(s) - \frac{\alpha'(s)^2}{4} + \frac{\alpha''(s)}{2};$$
$$\hat{q} = e^{-\frac{\alpha(s)}{2}}\tilde{q} \equiv \sqrt{\frac{u_o'}{u'(s)}} \cdot e^{\phi(s)} \cdot \tilde{\rho}(s,\vec{k}s) \; ; \; \varsigma(s) = e^{-\frac{\alpha(s)}{2}}(\tilde{q}_o''(s) - \tilde{q}_o'(s)\alpha'(s)). \quad (48)$$

It is known [44] that solution of homogeneous Hill's equation is represented by a symplectic matrix:

$$\begin{bmatrix} \hat{q}(s) \\ \hat{q}'(s) \end{bmatrix} = \mathbf{R}(s)\begin{bmatrix} \hat{q}(0) \\ \hat{q}'(0) \end{bmatrix}; \; \mathbf{R} = \begin{bmatrix} r_{11} & r_{12} \\ r_{21} & r_{22} \end{bmatrix}; \mathbf{R}' = \begin{bmatrix} 0 & 1 \\ -\hat{K}(s) & 0 \end{bmatrix}\mathbf{R}; \det\mathbf{R} = 1;, \quad (49)$$

which also defines general solutions of the inhomogeneous equation:

$$\hat{q}(s) = r_{11}(s)\hat{q}(0) + r_{12}(s)\hat{q}'(0) + \int_0^s (r_{11}(\zeta)r_{12}(s) - r_{11}(s)r_{12}(\zeta))\varsigma(\zeta)d\zeta. \quad (50)$$

Hence, solution of the homogenous Hill's equation $\hat{q}'' + \hat{K}(s)\hat{q} = 0$ can be used for investigation of this instability when the separation (45) is possible.

Cold beam, which is very popular in the studies of instabilities, has momenta distribution,

$$f_0(P) = \delta(P_1)\delta(P_2)\delta(P_3)$$

---

[11] Unfortunately, as can be seen from Eq.(44), such separation is impossible for Gaussian momenta distribution.

[12] It is easy to show that the separation $L_d(s,\zeta) = \phi_1(\zeta)\phi_2(s)$, with $\phi_2(s) \neq 0$ is also a sufficient condition. But we did not find cases, when such generalization is needed.

definitely satisfies this requirement with $\phi=1$. A more general and more interesting case is the beam with $\kappa$-1, also known as Lorentzian, momentum distribution in all three directions:

$$F_0(P) = F_{\kappa-1}(P) = \frac{1}{\pi^3}\prod_{i=1}^{3}\frac{\sigma_i}{\sigma_i^2 + P_i^2}, \quad (51)$$

allowing to integrate over the momenta:

$$L_d(\zeta,s) = \int e^{i(\vec{\eta}(\zeta)-\vec{\eta}(s))\cdot\vec{P}} F_o(P)dP^3 = e^{-\sum_{i=1}^{3}\sigma_i|\eta_i(s)-\eta_i(\zeta)|}. \quad (52)$$

If condition $|\eta_i(s)| \geq |\eta_i(\zeta)|$; $s \geq \zeta$ is satisfied for all three components $i=1,2,3$, we can use $|\eta_i(s)-\eta_i(\zeta)| = |\eta_i(s)|-|\eta_i(\zeta)|$ and the separated variables:

$$L_d(s,\zeta) = e^{\phi(\zeta)}e^{-\phi(s)}; \quad \phi(\zeta) = \sum_{i=1}^{3}\sigma_i|\eta_i(\zeta)|; \quad (53)$$

resulting in a second order ODE (50) for $\tilde{q}_{\vec{k}}(s) = \tilde{\rho}(s,\vec{k}(s))\cdot\exp(\phi(s))$, and in Hill's equation (48) for $\hat{q}(s) = \tilde{\rho}(s,\vec{k}(s))\sqrt{u'_o/u'(s)}\cdot\exp(\phi(s))$.

This is a good place to discuss driving term, $\varsigma(s)$, in the right-hand side (r.h.s.) of Hills equation

$$\varsigma(s) = e^{-\frac{\alpha}{2}}(\tilde{q}''_o - \tilde{q}'_o\alpha'); \quad \tilde{q}_o(s) = e^{\phi(s)}\tilde{\rho}_{\vec{k}}(s) = e^{\phi(s)}\int e^{-i\vec{\eta}(s)\cdot\vec{P}}\tilde{f}_{\vec{k}}(P)dP^3. \quad (54)$$

Generally speaking, for an arbitrary initial perturbation $\tilde{f}(q,P)$, both $\tilde{q}''_o$ and $\tilde{q}'_o$ are not equal zero and Hill's equation remains inhomogeneous. One case is an exception: when the initial perturbation is a product of the density perturbation and a κ-1 momenta distributions:

$$\tilde{f}(q,P) = \tilde{\rho}_o(q)\cdot f_{\kappa 1}(P) \rightarrow \int e^{-i\vec{\eta}(s)\cdot\vec{P}}f_{\kappa 1}(P) = \tilde{\rho}_o(q)\cdot e^{-\phi(s)};$$
$$\tilde{q}_o(s) = \tilde{\rho}_{\vec{k}} = \int \tilde{\rho}_o(q)e^{-i\vec{k}\cdot(q)}dq^3 = const, \quad (55)$$

all derivatives of $\tilde{q}_o$ are equal zero, and the Hill's equation becomes homogenous:

$$\tilde{f}(P,Q) = \tilde{\rho}_o(q)\cdot f_{\kappa 1}(P) \rightarrow \hat{q}'' + \hat{K}(s)\hat{q} = 0. \quad (56)$$

While the conditions $|\eta_i(s)| \geq |\eta_i(\zeta)|$; $s \geq \zeta$ are frequently satisfied, they also can be violated in the case of an arbitrary coupling. In fact, it is possible to construct matrix $\mathbf{U}$ that one component of vector $\vec{\eta} = \vec{k}\cdot\mathbf{U}$ turns from non-zero value at $\zeta$ to zero at $s > \zeta$. Emittance exchange lattices can serve as an example [42]. If even one of $|\eta_i(s)| \geq |\eta_i(\zeta)|$ conditions is violated, the separation becomes impossible and use of ODE is invalid. Nevertheless, the linear integral equation (40) is always solvable [43-44].

## IV. Special cases

We show in Appendix A (see Eq(A14)), that in the case of uncoupled motion the matrix $\mathbf{U}$ is diagonal with monotonically growing diagonal terms:

$$\mathbf{U}(s) = [\delta_{ij}]\mu_i(s); \; \mu_i(0) = 0; \; \mu_i(s) > \mu_i(\zeta) \; \forall \zeta < s \; \delta_{ij}\mu_i$$

which means that

$$|\eta_i(s)| = |k_{io}|\mu_i(s) \tag{57}$$

are also monotonic functions satisfying conditions $|\eta_i(s)| \geq |\eta_i(\zeta)|; s \geq \zeta$. Hence, we proved that in arbitrary accelerator with decoupled motion one can use second order ODE (47) or Hill's equation (48) for beam with $\kappa$-1 momentum (energy) distributions. This also includes linear accelerators using solenoids – the equations of motion can be easily decoupled by using torsion $\kappa_o(s) = -\dfrac{eB_s}{2p_oc}$ (see Eq.(1) and ref. [45]).

For the beam with the constant density and constant energy propagating in a drift space, all matrices and all components in the equations can be easily evaluated:

$$\mathbf{A} = \mathbf{D} = \mathbf{I}; \; \mathbf{C} = \mathbf{0}; \; \mathbf{U}(s) = \mathbf{B}(s) = \frac{1}{\gamma_o\beta_o mc}\begin{bmatrix} s & 0 & 0 \\ 0 & s & 0 \\ 0 & 0 & s/\gamma_o^2\beta_o^2 \end{bmatrix}; \tag{58}$$

$$\vec{k} = const; \; \tilde{\rho}_{\vec{k}} = const; \; u = \frac{s}{\gamma_o^3\beta_o^3 mc}\left(\gamma_o^2\beta_o^2\vec{k}_\perp^2 + k_3^2\right); \; u'K = \frac{4\pi n_o e^2}{\gamma_o^3\beta_o^3 mc} = const; \; u'' = 0.$$

For cold plasma oscillations it results in $\vec{k}$-independent equation:

$$\frac{d^2\tilde{\rho}_{\vec{k}}}{ds^2} + k_p^2\tilde{\rho}_{\vec{k}} = 0; \; k_p^2 = \frac{4\pi n_o e^2}{\gamma_o^3\beta_o^3 mc}, \tag{59}$$

which, after applying the inverse Fourier transformation, becomes the carbon copy of known plasma oscillations but in the laboratory frame:

$$\frac{d^2\tilde{\rho}(\vec{q})}{ds^2} + k_p^2\tilde{\rho}(\vec{q}) = 0; \; \tilde{\rho}(\vec{q}) = \tilde{\rho}_o(\vec{q})e^{i(k_p s - \omega_{pl} t)}; \; \omega_{pl} = c\beta_o k_p. \tag{60}$$

For beams with $\kappa$-1 momentum distribution (51) propagating in a drift space with the constant density and constant energy, $\vec{k}$-dependence occurs via Landau damping term and $q''_{\vec{k}}$ in the driving term:

$$\tilde{q}'' + k_p^2(s)\tilde{q} = \tilde{q}_o''; \quad \tilde{q} = e^{\phi(s)}\tilde{\rho}_{\vec{k}}; \tilde{q}_o = e^{\phi(s)}\tilde{\rho}_{\vec{k}};$$

$$\phi(s) = \frac{s}{\gamma_o \beta_o mc} \cdot \left(\sigma_1|k_1| + \sigma_2|k_2| + (\gamma_o \beta_o)^{-2}\sigma_3|k_3|\right) \quad (61)$$

$$\tilde{\rho}_{\vec{k}} = \int \exp\left(i\frac{s}{\gamma_o \beta_o mc} \cdot \left(k_1\sigma_1 \underline{P}_1 + \sigma_2 k_2 \underline{P}_2 + (\gamma_o \beta_o)^{-2}\sigma_3 k_3 \underline{P}_3\right)\right) \underline{\tilde{f}}_{\vec{k}}(\underline{P}) d\underline{P}^3.$$

One important consequence of Eq.(61) is that for beam propagating in straight section, the Landau damping decrement for transverse modulation is boosted by factor $\gamma_o^2 \beta_o^2$, which is typically $\gg 1$. In other words, for $|k_{1,2}|\sigma_{1,2} \propto |k_3|\sigma_3$, the Landau damping term is significantly larger than for $k_{1,2} = 0$. This is one of the reasons why longitudinal instability plays a special role and is of a special interest for accelerators with $\gamma_o^2 \gg 1$.

Let's consider 1D longitudinal instability in a beam propagating along straight trajectory, e.g. when the longitudinal and transverse motion are decoupled:

$$\mathbf{A}(s) = \begin{bmatrix} \mathbf{A}_\perp(s) & 0 \\ 0 & a_\parallel(s) \end{bmatrix}; \quad \mathbf{B}(s) = \begin{bmatrix} \mathbf{B}_\perp(s) & 0 \\ 0 & b_\parallel(s) \end{bmatrix}; \quad \vec{k}(s) = \hat{e}_3 k(s) = \hat{e}_3 \frac{k_o}{a_\parallel(s)}. \quad (64)$$

Evolution of this instability can be described either by integral equation:

$$\tilde{\rho}(s, k(s)) = -\frac{4\pi n_o e^2}{c} \int_o^s \tilde{\rho}(\zeta, k(\zeta)) K_\parallel(\zeta) (u(s) - u(\zeta)) d\zeta \int e^{-ik_o(u(s)-u(\zeta))\cdot \underline{P}} \underline{f}_\parallel(\underline{P}) d\underline{P} + \tilde{\rho}_k(s);$$

$$K_\parallel(\zeta) = \frac{4\pi n_o e^2}{c} \frac{a_\parallel(\zeta)}{\det \mathbf{A}_\perp(\zeta)}; \quad u(\zeta) = \frac{b_\parallel(\zeta)}{a_\parallel(\zeta)}; \quad \tilde{\rho}_{ok}(s) = \int e^{-ik_o u(s)\underline{P}} \underline{\tilde{f}}_{k'\parallel}(\underline{P}) d\underline{P}.$$

(63)

or for κ-1 longitudinal momentum distribution by differential equation:

$$\tilde{q}'' - \xi(s)\cdot \tilde{q}' + k_p^2(s)\tilde{q} = \tilde{q}_o'' - \xi(s)\cdot \tilde{q}_o';$$

$$k_p^2(s) = \frac{4\pi}{(\gamma_o \beta_o)^3} \cdot \frac{n_o r_c}{a_\parallel \det \mathbf{A}_\perp}; \xi(s) = \frac{d}{ds}\left(\ln a_\parallel^2 (\gamma_o \beta_o)^3\right); \tilde{q}(s) = \tilde{\rho}\left(s, \frac{k_o}{a_\parallel(s)}\right) e^{\frac{k_o b_\parallel(s)\sigma_3}{a_\parallel(s)}}, \quad (64)$$

where $k_p(s)$ is $s$-depended plasma "frequency" (in $s$ domain), $k/a_\parallel(s)$ is scaled wavenumber of the perturbation, and $\xi(s)$ represents an addition term, which, depending on its sign, either damps or amplifies modulation. The corresponding Hill's equation has the same driving term but slightly different $s$-depended "frequency":

$$\hat{q}'' + k_p'^2 \hat{q} = a_\parallel (\gamma_o \beta_o)^{3/2}\left(\tilde{q}_o'' - \tilde{q}_o' \frac{u''}{u'}\right); \quad k_p'^2 = k_p^2 - \frac{\xi'^2}{4} + \frac{\xi''}{2}. \quad (65)$$

For beam with the constant energy and no-compression ($a_\parallel = 1$), equations (64) and (65) become identical. They describe the instability driven by transverse focusing:

$$a_\parallel = 1;\ \gamma_o\beta_o = const; b_\parallel = \frac{s}{(\gamma_o\beta_o)^3 mc} \to \tilde{q}'' + k_p^2(s)\tilde{q} = \tilde{q}_o'';$$

$$k_p^2(s) = \frac{4\pi}{(\gamma_o\beta_o)^3} \cdot \frac{n_o r_c}{\det \mathbf{A}_\perp(s)};\ \rho_k(s) = \tilde{q}(s) e^{-\frac{ks}{(\gamma_o\beta_o)^2}\frac{\sigma_{\gamma_o}}{\gamma_o\beta_o}}.$$

(66)

### V. Discussions

To our knowledge, the analytical solution for the evolution of the finite size beams that are strongly affected by the SC forces has never been found before. In this paper, we presented the most general theoretical description of 3D microscopic (short wavelength) instabilities driven by the SC forces. Our approach of solving this problem is based on the local linearization of the nonlinear transfer map, which includes the macroscopic SC forces. This approach allows to arrive to a linear Vlasov equation for the microscopic perturbations at the scale much smaller than the other important scales of the problem (beam sizes, changes in the beam trajectory, scales of nonlinearity in the transfer map, etc. – see Eq. (18-20).

As was suggested in the previous section, matrix $\mathbf{A}$ plays a special role in the evolution of the microscopic perturbation in beams with strong SC. First and foremost, approximation of the homogenous background results in the beam density, $n(s)$, to be inversely proportional to determinant of matrix $\mathbf{A}$: $n(s) = n_o \det \mathbf{A}^{-1}$. It is also easy to show that this is no longer a problem for a beam with finite sizes and finite emittances where beam density remains finite when $\det \mathbf{A} = 0$ [13]. It means, that in the final expression for instability, we can use calculated finite local density $n(s)$:

$$\tilde{\rho}(s, \vec{k}(s)) = -\int_o^s \frac{4\pi n(\zeta) e^2}{cv(\zeta)} \tilde{\rho}(\zeta, \vec{k}(\zeta))(u(s) - u(\zeta)) L_d(s, \zeta) d\zeta + \tilde{\rho}_{\vec{k}_o}(s);$$

(67)

The second complication related to $\det \mathbf{A} \to 0$ arises with $k$-vector transformation (29):

$$\underline{k}^T(\zeta) = \underline{k}^T \cdot \mathbf{A}^{-1}(\zeta),$$

---

[13] For Gaussian distribution, $m_{11}=0$ simply means rotation by 90 degrees in the phase space and the momentum spread determines the density:

$$f_o = \frac{1}{2\pi\sigma_q\sigma_P}\exp\left(-\frac{q^2}{2\sigma_q^2} - \frac{P^2}{2\sigma_P^2}\right) = \frac{1}{2\pi\sigma_q\sigma_P}\exp\left(-\frac{(m_{22}q - m_{12}P)^2}{2\sigma_Q^2} - \frac{(-m_{21}q + m_{11}P)^2}{2\sigma_P^2}\right);$$

$$\rho(q,s) = \int_{-\infty}^{\infty} f_o\, d\underline{P} = \frac{1}{\sqrt{2\pi(m_{11}^2\sigma_q^2 + m_{12}^2\sigma_P^2)}}\exp\left(-\frac{q^2}{2(m_{11}^2\sigma_q^2 + m_{12}^2\sigma_P^2)}\right).$$

which involves inversion of matrix $\mathbf{A}$. In contrast with the beam density, which remains finite for a finite size beams, the module of *k*-vector is generally not limited: $|k| \underset{\det \mathbf{A} \to 0}{\to} \infty$.

The resolution of this challenge is that the kernel in the integral remains finite:

$$\left|\frac{u(s)-u(\zeta)}{\upsilon(\zeta)}\right| \propto \frac{|k(s)|}{|k(\zeta)|^2} + \frac{|k(\zeta)|}{|k(\zeta)|^2} \propto \frac{\det \mathbf{A}(\zeta)^2}{\det \mathbf{A}(s)} + \det \mathbf{A}(\zeta).$$

For any finite momenta spreads and $|\mathbf{B}| \neq 0$, the Landau damping term vanishes when $|k| \to \infty$:

$$L_d = \int e^{i(\vec{k}(\zeta)\cdot\tilde{\mathbf{B}}(\zeta)-\vec{k}(s)\cdot\tilde{\mathbf{B}}(s))\cdot\vec{P}} \underline{f}_o(P) dP^3 \underset{|\vec{k}(\zeta)\tilde{\mathbf{B}}(\zeta)|+|\vec{k}(s)\tilde{\mathbf{B}}(s)|\to\infty}{\Rightarrow} 0,$$

which resolves uncertainties how critical compression points can be evaluated for an arbitrary initial perturbation in density momenta. With known – to be exact, properly evaluated in specific portion of phase space – transport matrices, one can numerically solve equations (40) or (67) – see example in **Appendix E**.

An additional complication of using the method presented in this paper can arise when simultaneous compliance with conditions (18) and (20) is impossible. In other words, an accelerator lattice with strong filamentation of phase space caused by nonlinearity may not allow to define a phase space area where linearization of the map (20) and conditions for homogenous background density (18) are compatible. For example, a multi-million turn map with strong amplitude dependent tune shifts could epitomize such case.

It is desirable to include CSR in the evolution of the microscopic density evolution, especially for high energy accelerators with strong bends. Unfortunately, CSR perturbations are not local, and application of procedure used in this paper could be erroneous.

## VI. Conclusions and Acknowledgements.

In this paper we derived linear integral equation uniquely describing evolution of 3D microscopic density modulation driven by SC forces, including instabilities. Our theory includes the coupling between density modulation in various degrees of freedom, for example occurring in a bend, in a SQ-quadrupole or in a transverse deflecting cavity. It also includes effects of compression in all three directions, rotation, energy chirp and acceleration (deceleration) of the beams as well as Landau damping. Most likely it will be most useful for investigating beam stability in low energy linear accelerators with SC dominated beam.

We also derived conditions when linear integral equation can be reduced to ordinary second order differential equation and demonstrated application of our method for a set of special cases.

This research was supported by NSF grant PHY-1415252, by DOE NP office Award DE- FOA-0000632, DOE HEP Award DE-SC0020375, and by Brookhaven Science Associates, LLC under Contract No. DE-SC0012704 with the U.S. Department of Energy.

.

**Appendix A – System Hamiltonian and equations of motion**

Traditionally in accelerator physics literature the phase space vector is combined from Canonical pairs of coordinates and momenta $\left[...(q_i, P^i)...\right] \equiv \left[...(x_{2i-1}, x_{2i})...\right]$. In this paper, following A. Dragt [34], we use equivalent, but different structure of the *phase-space* vector, which clearly separate coordinates and momenta and simplifies form of the matrix of symplectic generator, **S**;

$$\xi^T = [\xi_1,...,\xi_{2n}] = [q^T, P^T]; q^T = [q_1,...,q_n]; P^T = [P^1,...,P^n];$$

$$\begin{cases} \dfrac{dq_i}{ds} = \dfrac{\partial H}{\partial P^i} \\ \dfrac{dP^i}{ds} = -\dfrac{\partial H}{\partial q_i} \end{cases} \Leftrightarrow \dfrac{d\xi}{dt} = \mathbf{S}\dfrac{dH}{d\xi} \Leftrightarrow \dfrac{d\xi_i}{dt} = \mathbf{S}_{ij}\dfrac{dH}{d\xi_j}; \quad (A1)$$

$$\mathbf{S} \equiv [S_{ik}] = \begin{bmatrix} \mathbf{0} & \mathbf{I} \\ -\mathbf{I} & \mathbf{0} \end{bmatrix}; \mathbf{I}_{n\times n} = \begin{bmatrix} 1 & 0 & 0 \\ 0 & ... & 0 \\ 0 & 0 & 1 \end{bmatrix}.$$

Use of these notations is especially convenient for linear maps in the form of *2nx2n* symplectic transport matrices:

$$\xi(s_2) \equiv \begin{bmatrix} q(s_2) \\ P(s_2) \end{bmatrix} = \mathbf{M}(s_1|s_2)\xi(s_1) \equiv \mathbf{M}(s_1|s_2)\begin{bmatrix} q(s_1) \\ P(s_1) \end{bmatrix};$$

$$\mathbf{M}^T\mathbf{S}\mathbf{M} = \mathbf{M}\mathbf{S}\mathbf{M}^T = \mathbf{S}; \mathbf{M}^{-1} = -\mathbf{S}\mathbf{M}^T\mathbf{S};$$

$$\mathbf{M} = \begin{bmatrix} \mathbf{A} & \mathbf{B} \\ \mathbf{C} & \mathbf{D} \end{bmatrix}; \mathbf{M}^{-1} = \begin{bmatrix} \mathbf{D}^T & -\mathbf{B}^T \\ -\mathbf{C}^T & \mathbf{A}^T \end{bmatrix}; \quad (A2)$$

$$q(s_2) = \mathbf{A}q(s_1) + \mathbf{B}P(s_1); P(s_2) = \mathbf{C}q(s_1) + \mathbf{D}P(s_1);$$

$$q(s_1) = \mathbf{D}^T q(s_2) - \mathbf{B}^T P(s_2); P(s_1) = -\mathbf{C}^T q(s_2) + \mathbf{A}^T P(s_2);$$

providing explicit connection between coordinates and momenta with their initial values and vice versa. It also provides important properties of the block matrices which can be very useful for the evaluation of complex expression. Specifically, symplecticity of transport matrix requires that four *nxn* matrices $\mathbf{AB}^T, \mathbf{DC}^T, \mathbf{A}^T\mathbf{C}, \mathbf{D}^T\mathbf{B}$ will be symmetric

$$\left(\mathbf{AB}^T\right)^T = \mathbf{AB}^T; \left(\mathbf{DC}^T\right)^T = \mathbf{DC}^T, \left(\mathbf{A}^T\mathbf{C}\right)^T = \mathbf{A}^T\mathbf{C}, \left(\mathbf{D}^T\mathbf{B}\right)^T = \mathbf{D}^T\mathbf{B} \quad (A3)$$

and that

$$\mathbf{AD}^T - \mathbf{BC}^T = \mathbf{I}; \quad \mathbf{A}^T\mathbf{D} - \mathbf{C}^T\mathbf{B} = \mathbf{I}. \quad (A4)$$

In the case of uncoupled motion, all four matrices $\mathbf{A}, \mathbf{B}, \mathbf{C}, \mathbf{D}$ become diagonal automatically satisfying conditions (A3) and turning (A4) into simpler conditions for diagonal components of block-matrices:

$$A_{ii}D_{ii} - B_{ii}C_{ii} = 1; \quad i = 1,..,n \tag{A5}$$

equivalent to unity of determinants for individual *2x2* matrices in notations (A2):

$$M_i = \begin{bmatrix} A_{ii} & B_{ii} \\ C_{ii} & D_{ii} \end{bmatrix}; \det M_i = 1. \tag{A6}$$

Matrix $\mathbf{A}^{-1}\mathbf{B}$ plays critical role in the instability integral equation (40). Its properties can be studied for a Hamiltonian system describing a generic linear system:

$$H = \frac{1}{2}\xi^T \mathbf{H}(s)\xi; \ \mathbf{H}^T = \mathbf{H} = \begin{bmatrix} \mathbf{H}_q & \mathbf{H}_m^T \\ \mathbf{H}_m & \mathbf{H}_p \end{bmatrix}; \mathbf{H}_{q,p}^{\ T} = \mathbf{H}_{q,p}. \tag{A7}$$

Using equations of motion, the derivative of the matrix M becomes:

$$\mathbf{M}' \equiv \frac{d\mathbf{M}}{ds} = \mathbf{SH}\cdot\mathbf{M}; \ \mathbf{A}' = \mathbf{H}_m\mathbf{A} + \mathbf{H}_p\mathbf{C}; \ \mathbf{B}' = \mathbf{H}_m\mathbf{B} + \mathbf{H}_p\mathbf{D}. \tag{A8}$$

Taking into account that $\left(\mathbf{A}^{-1}\right)' = -\mathbf{A}^{-1}\mathbf{A}'\mathbf{A}^{-1}$, we get

$$\left(\mathbf{A}^{-1}\mathbf{B}\right)' = \mathbf{A}^{-1}\mathbf{B}' - \mathbf{A}^{-1}\mathbf{A}'\mathbf{A}^{-1}\mathbf{B} = \mathbf{A}^{-1}\mathbf{H}_p\left(\mathbf{D} - \mathbf{C}\mathbf{A}^{-1}\mathbf{B}\right), \tag{A9}$$

which can be turned into

$$\begin{aligned} \left(\mathbf{A}^{-1}\mathbf{B}\right)' &= \mathbf{A}^{-1}\mathbf{H}_p\left(\mathbf{A}^T\right)^{-1} \\ \mathbf{D} - \mathbf{C}\mathbf{A}^{-1}\mathbf{B} &= \left(\mathbf{D}\mathbf{A}^T - \mathbf{C}\mathbf{B}^T\right)\left(\mathbf{A}^{-1}\right)^T = \left(\mathbf{A}^{-1}\right)^T; \\ \left(\mathbf{A}^{-1}\mathbf{B}\right)' &= \mathbf{A}^{-1}\mathbf{H}_p\left(\mathbf{A}^{-1}\right)^T \end{aligned} \tag{A10}$$

using symplecticity conditions (A4) $\mathbf{A}^{-1}\mathbf{B} = \mathbf{B}^T\left(\mathbf{A}^T\right)^{-1}$ and $\mathbf{D}\mathbf{A}^T - \mathbf{C}\mathbf{B}^T = \mathbf{I}$, one can show that

$$\mathbf{D} - \mathbf{C}\mathbf{A}^{-1}\mathbf{B} = \mathbf{D} - \mathbf{C}\mathbf{B}^T\left(\mathbf{A}^T\right)^{-1} = \left(\mathbf{D}\mathbf{A}^T - \mathbf{C}\mathbf{B}^T\right)\left(\mathbf{A}^T\right)^{-1} = \left(\mathbf{A}^T\right)^{-1}.$$

It is possible to show for an arbitrary accelerator [34-36] that $\mathbf{H}_p$ is a diagonal with positive diagonal terms:

$$\mathbf{H}_p = \frac{1}{\gamma_o \beta_o mc} \begin{bmatrix} 1 & 0 & 0 \\ 0 & 1 & 0 \\ 0 & 0 & (\gamma_o \beta_o)^{-2} \end{bmatrix} \tag{A11}$$

This allows us to prove that for an arbitrary accelerator with invertible matrix $\mathbf{A}$, the convolution $u(s)$ in Eq.(40) is nonnegative monotonically growing function with $u'(s) > 0$.

$$u(\zeta) = \vec{k}(\zeta) \cdot \dot{\mathbf{B}}(\zeta) \cdot \vec{k} \equiv \vec{k}(\zeta) \cdot \dot{\mathbf{U}}(\zeta) \cdot \vec{k}$$

$$u(s) = \underline{k}^T \left( \mathbf{A}(s)^{-1} \mathbf{B}(s) \right) \underline{k} \equiv \vec{k} \overrightarrow{\left( \mathbf{A}^{-1} \mathbf{B} \right)} \vec{k}; \ \mathbf{B}(0) = \mathbf{0} \rightarrow u(0) = 0; \ \vec{k}(s) = \mathbf{A}^T(s)^{-1} \vec{k};$$

$$u'(s) = \underline{k}^T \left( \mathbf{A}^{-1} \mathbf{B} \right)' \underline{k} = k^T(s) \mathbf{H}_p(s) k(s) = \sum_{i=1}^{3} \mathbf{H}_{ii}(s) k_i^2(s) > 0. \tag{A12}$$

$$u(s) = \int_0^s \left( \sum_{i=1}^{3} \mathbf{H}_{ii}(s) k_i^2(s) \right) d\zeta \geq 0.$$

In other words, we proved that convolution of any constant vector with matrix $\mathbf{A}^{-1}\mathbf{B}$ is nonnegative monotonically growing function.

In the chase of uncoupled motion, all 3x3 matrices are diagonal and

$$\frac{d}{ds}\left(\mathbf{A}^{-1}\mathbf{B}\right) = \begin{bmatrix} \alpha_1(s) & 0 & 0 \\ 0 & \alpha_2(s) & 0 \\ 0 & 0 & \alpha_3(s) \end{bmatrix} = \frac{1}{\gamma_o \beta_o mc} \begin{bmatrix} a_{11}^{-2} & 0 & 0 \\ 0 & a_{22}^{-2} & 0 \\ 0 & 0 & \left(a_{33}\gamma_o\beta_o\right)^{-2} \end{bmatrix}; \alpha_i(s) > 0;$$

$$\mu_i(s) = \int_0^s \alpha_i(\zeta) d\zeta \geq 0; \ \mathbf{A}^{-1}\mathbf{B} = \int_0^s \left(\mathbf{A}^{-1}\mathbf{B}\right)' d\zeta = \begin{bmatrix} \mu_1(s) & 0 & 0 \\ 0 & \mu_2(s) & 0 \\ 0 & 0 & \mu_3(s) \end{bmatrix} = \delta_{ij}\mu_i. \tag{A13}$$

i.e. diagonal terms of matrix $\mathbf{A}^{-1}\mathbf{B}$ are monotonically growing positive functions:

$$\forall s_1 > s_2; \mu_i(s_1) > \mu_i(s_2). \tag{A14}$$

**Appendix B – Conditions for applicability of the short period (microscopic) perturbations.**

These conditions are also known as assumptions of homogeneous infinite plasma. Fourier or Laplace transformations are frequently used to solve the linearized Vlasov equation. The main problem from inhomogeneous distribution (or finite size of the beam) is that it results in coupling between the Fourier harmonics of the perturbation and those of the background, e.g., applying Fourier transformation to Eq. (11)

$$\int dQ^3 e^{-i\vec{k}\vec{Q}} \left( \frac{\partial \tilde{f}}{\partial s} + \frac{\partial f_o}{\partial \vec{q}} \frac{\partial \tilde{h}}{\partial \vec{P}} - \frac{\partial f}{\partial \vec{P}} \frac{\partial \tilde{h}}{\partial \vec{q}} \right) =$$

$$\frac{\partial \tilde{f}_{\vec{k}}}{\partial t} + i \int d\mathbf{k}^3 \left\{ f_{o\mathbf{k}} \left( \vec{k} \cdot \frac{\partial \tilde{h}_{\vec{k}-\mathbf{k}}}{\partial \vec{P}} \right) - \tilde{h}_{\vec{k}-\mathbf{k}} \left( (\vec{k}-\mathbf{k}) \cdot \frac{\partial f_{o\mathbf{k}}}{\partial \vec{P}} \right) \right\},$$

does not result in separation of the Fourier harmonics. In this sense, this equation is as complicated as the original Vlasov equation. The conditions for separation of the Fourier harmonics are easiest to derive in the comoving frame of reference. In this Appendix we will use indexes "*cm*" and "*lab*" to distinguish between the comoving and laboratory frames, correspondingly.

In a vicinity of azimuth $s_o$, particle's trajectory in the laboratory frame can be described using Cartesian coordinate system with three fixed orthogonal unit vectors $\hat{e}_{1,2,3}(s_o)$ (see Section II):

$$\vec{r}_{lab} = \vec{r}_o(s_o) + \hat{e}_1(s_o)x + \hat{e}_2(s_o)y + \hat{e}_3(s_o)z.$$

Let's consider an instantaneous co-moving frame that propagates in vicinity of azimuth $s_o$ with velocity

$$\vec{v}_o = v_o(s_o) \cdot \hat{e}_3(s_o),$$

along local *z*-axis. Next step is to establish relations between the parameters in the laboratory and the comoving frame. It is known that Lorentz transformation does not affect transverse coordinates $x, y$, but boosts longitudinal coordinate by the relativistic factor $\gamma_o = \left(1 - \frac{\vec{v}_o^2}{c^2}\right)^{-1/2}$:

$$z_{cm} = \gamma_o z \Rightarrow \vec{r}_{cm} = \hat{x} \cdot x + \hat{y} \cdot y + \gamma_o \hat{z} \cdot z;$$

and also transforms the 4-vectors $\hat{k} = (\omega/c, \vec{k})$ as [37]:

$$\vec{k}_{cm} \equiv \vec{k} = \hat{z}k_z + \vec{k}_\perp; \omega_{cm} = 0; \vec{k}_{lab} \equiv \vec{k} = \hat{z}k_z + \vec{k}_\perp;$$

$$\vec{k}_\perp = \vec{k}_\perp; k_z = \gamma_o \left( k_z + \beta_o \frac{\omega_{cm}}{c} \right) = \gamma_o k_z; \omega_{lab} = \gamma_o (\omega_{cm} + v_o k_z) = v_o k_z; \quad (B1)$$

with known relation between exponents in Fourier transforms:

$$e^{i\vec{k}\vec{r}_{cm}} = e^{i(\vec{k}\vec{r}_{lab} - v_o k_z (t - t_o(s_o)))} = e^{i\left(\vec{k}\vec{r}_{lab} + \frac{k_3 q_3}{\beta_o(s_o)}\right)}$$

providing us with important connection with k-vector defined in Eq.(18) of the main text:

$$k_1 = k_x = \mathbf{k}_x;\ k_2 = k_y = \mathbf{k}_y;\ k_3 = \beta_o k_z = \gamma_o \beta_o \mathbf{k}_z;$$
$$a_1 = a_x, a_2 = a_y; a_3 = a_z / \beta_o.$$
(B2)

Let's consider a beam with typical scales of the inhomogeneity, $a_{x,y,z}$, which are not necessarily of the same order of magnitude:

$$\left|\frac{\partial f_o}{\partial x}\right| \propto \frac{f_o}{a_x}; \left|\frac{\partial f_o}{\partial y}\right| \propto \frac{f_o}{a_y}; \left|\frac{\partial f_o}{\partial z}\right| \propto \frac{f_o}{a_z}$$

defined in the laboratory frame. As indicated above, transverse scales will remain the same in the co-moving frame, but longitudinal scale will be boosted by factor $\gamma_o$.

For simplicity, we will consider that the particle motion in the comoving frame is non-relativistic, and we can neglect effects of the magnetic field, e.g., assume $\vec{B}_{cm} = 0$. In this case Maxwell equations are reduced to two equations for electric field:

$$div\vec{E} = 4\pi\rho;\ curl\vec{E} = 0.$$
(B3)

Further in this Appendix we will use the comoving frame and will drop the index "*cm*". The natural condition for neglecting the beam's edges, transitions and reflection effects is that there must be a significant number of oscillations in each direction at the typical scales of the inhomogeneity, e.g.:

$$\mathbf{k}_{x,y} a_{x,y} = k_{x,y} a_{x,y} \gg 2\pi;\ \gamma \mathbf{k}_z a_z = k_z a_z \gg 2\pi.$$
(B4)

But as we find out in this Appendix, not all of this "natural" conditions are necessary. For example, it is intuitively understandable that for one-dimensional perturbation with $\mathbf{k}_z \neq 0, \mathbf{k}_{x,y} = 0$, two requirments in (B4), $\mathbf{k}_{x,y} a_{x,y} \gg 2\pi$ are not necessary. As we will also show that condition (B4) in $i^{th}$ direction is not necessary if $|\mathbf{k}_i| \ll |\vec{\mathbf{k}}|$.

Second, and much more convoluted, condition is that Fourier harmonic of the induced electric field (and therefore of the perturbation in the Hamiltonian) are linear functions of the harmonic of the charge density perturbation,

$$\rho_{\vec{k}} = e\int_{-\infty}^{\infty} d\vec{r} e^{-i\vec{k}\vec{r}} \int_{-\infty}^{\infty} \tilde{f}_{cm}(\vec{r}, \vec{v}, t) d\vec{v}.$$
(B5)

where $\tilde{f}_{cm}$ is a perturbation of the distribution function in the comoving frame. In an infinite charged plasma, a periodic density perturbation results in a periodic electric field aligned with the $\vec{\mathbf{k}}$-vector:

$$\vec{E} = \vec{E}_{\vec{k}} e^{i\vec{k}\vec{r}};\ \vec{E}_{\vec{k}} = \vec{E}_{\vec{k}_\parallel} + \vec{E}_{\vec{k}_\perp};$$
$$curl\vec{E} = 0 \to \vec{\mathbf{k}} \times \vec{E}_{\vec{k}} = 0 \to \vec{E}_{\vec{k}_\perp} = 0 \to \vec{E}_{\vec{k}_\parallel} = \frac{\vec{\mathbf{k}}}{|\vec{\mathbf{k}}|} E_{\vec{k}};$$
(B6)
$$div\vec{E} = 4\pi\rho_{\vec{k}} e^{i\vec{k}\vec{r}} \to i\vec{\mathbf{k}} \cdot \vec{E}_{\vec{k}} = i|\vec{\mathbf{k}}| E_{\vec{k}} = 4\pi\rho_{\vec{k}},$$

resulting in

$$\vec{E}_{\vec{k}} = -4\pi i \rho_{\vec{k}} \frac{\vec{k}}{\vec{k}^2}. \tag{B7}$$

For a non-uniform density, electric field will deviate from intuitive extensions of (B7) by some $\delta \vec{E}$:

$$\vec{E} = 4\pi \rho_{\vec{k}}(\vec{r}) \frac{\vec{k}}{i\vec{k}^2} e^{i\vec{k}\vec{r}} + \delta \vec{E}. \tag{B8}$$

We can neglect $\delta \vec{E}$ when compared with the main r.h.s. term in (B7) when $|\vec{k}||\delta \vec{E}| \ll 4\pi|\rho_{\vec{k}}|$. We get the following using (B3):

$$\begin{aligned}
div\vec{E} &= 4\pi \rho_{\vec{k}}(\vec{r})e^{i\vec{k}\vec{r}} + 4\pi \frac{\left(\vec{k}\cdot\vec{\nabla}\rho_{\vec{k}}(\vec{r})\right)}{\vec{k}^2} e^{i\vec{k}\vec{r}} + div\delta\vec{E} = 4\pi \rho_{\vec{k}}(\vec{r})e^{i\vec{k}\vec{r}}; \\
curl\vec{E} &= 4\pi \frac{\left(\vec{k}\times\vec{\nabla}\rho_{\vec{k}}(\vec{r})\right)}{\vec{k}^2} e^{i\vec{k}\vec{r}} + curl\delta\vec{E} = 0; \\
div\delta\vec{E} &= -4\pi \frac{\left(\vec{k}\cdot\vec{\nabla}\rho_{\vec{k}}(\vec{r})\right)}{\vec{k}^2} e^{i\vec{k}\vec{r}} \sim |\vec{k}|\cdot|\delta\vec{E}| \\
curl\delta\vec{E} &= -4\pi \frac{\left(\vec{k}\times\vec{\nabla}\rho_{\vec{k}}(\vec{r})\right)}{\vec{k}^2} e^{i\vec{k}\vec{r}} \sim |\vec{k}|\cdot|\delta\vec{E}|
\end{aligned} \tag{B9}$$

While the error estimation resulting from $div\vec{E} = 4\pi \rho_{\vec{k}}$ improves on the intuitive requirement (B4):

$$\frac{\partial \delta E_x}{\partial x} + \frac{\partial \delta E_y}{\partial y} + \frac{\partial \delta E_z}{\partial z} = -4\pi \frac{e^{i\vec{k}\vec{r}}}{\vec{k}^2}\left(\mathbf{k}_x \cdot \frac{\partial \rho_{\vec{k}}}{\partial x} + \mathbf{k}_y \cdot \frac{\partial \rho_{\vec{k}}}{\partial y} + \mathbf{k}_z \cdot \frac{\partial \rho_{\vec{k}}}{\partial y}\right); \left|\frac{\partial \rho_{\vec{k}}}{\partial x_i}\right| \sim \frac{|\rho_{\vec{k}}|}{a_i};$$

$$\left|-ie^{i\vec{k}\vec{r}}\left(\frac{\partial \delta E_x}{\partial x} + \frac{\partial \delta E_y}{\partial y} + \frac{\partial \delta E_z}{\partial z}\right)\right| \sim \left(|\mathbf{k}_x \delta E_x| + |\mathbf{k}_y \delta E_y| + |\mathbf{k}_z \delta E_y|\right) \sim 4\pi \frac{|\rho_{\vec{k}}|}{k^2}\left(\frac{|\mathbf{k}_x|}{a_x} + \frac{|\mathbf{k}_y|}{a_y} + \frac{|\mathbf{k}_z|}{\gamma a_z}\right) \tag{B10}$$

$$\left|\frac{div\delta\vec{E}}{div\vec{E}}\right| \sim \frac{\frac{|\mathbf{k}_x|}{a_x} + \frac{|\mathbf{k}_y|}{a_y} + \frac{|\mathbf{k}_z|}{\gamma a_z}}{\mathbf{k}^2} \ll 1$$

the error estimations resulting from $curl\vec{E} = 0$

$$curl\delta\vec{E} = -4\pi \frac{\left(\vec{\mathbf{k}} \times \vec{\nabla} \rho_{\vec{\mathbf{k}}}(\vec{r})\right)}{\vec{\mathbf{k}}^2} e^{i\vec{\mathbf{k}}\vec{r}} = \frac{4\pi}{\vec{\mathbf{k}}^2} e^{i\vec{\mathbf{k}}\vec{r}} \times$$

$$\left\{ \hat{x}\left(\mathbf{k}_y \frac{\partial \rho_{\vec{\mathbf{k}}}}{\partial z} - \mathbf{k}_z \frac{\partial \rho_{\vec{\mathbf{k}}}}{\partial y}\right) + \hat{y}\left(\mathbf{k}_z \frac{\partial \rho_{\vec{\mathbf{k}}}}{\partial x} - \mathbf{k}_x \frac{\partial \rho_{\vec{\mathbf{k}}}}{\partial z}\right) + \hat{z}\left(\mathbf{k}_x \frac{\partial \rho_{\vec{\mathbf{k}}}}{\partial y} - \mathbf{k}_y \frac{\partial \rho_{\vec{\mathbf{k}}}}{\partial x}\right)\right\}$$

$$\frac{\partial \delta E_z}{\partial y} - \frac{\partial \delta E_y}{\partial z} = \frac{4\pi}{\vec{\mathbf{k}}^2} e^{i\vec{\mathbf{k}}\vec{r}} \left(\mathbf{k}_y \frac{\partial \rho_{\vec{\mathbf{k}}}}{\partial z} - \mathbf{k}_z \frac{\partial \rho_{\vec{\mathbf{k}}}}{\partial y}\right); \quad \frac{\partial \delta E_x}{\partial z} - \frac{\partial \delta E_z}{\partial x} = \frac{4\pi}{\vec{\mathbf{k}}^2} e^{i\vec{\mathbf{k}}\vec{r}} \left(\mathbf{k}_z \frac{\partial \rho_{\vec{\mathbf{k}}}}{\partial x} - \mathbf{k}_x \frac{\partial \rho_{\vec{\mathbf{k}}}}{\partial z}\right);$$

$$\frac{\partial \delta E_y}{\partial x} - \frac{\partial \delta E_x}{\partial y} = \frac{4\pi}{\vec{\mathbf{k}}^2} e^{i\vec{\mathbf{k}}\vec{r}} \left(\mathbf{k}_x \frac{\partial \rho_{\vec{\mathbf{k}}}}{\partial y} - \mathbf{k}_y \frac{\partial \rho_{\vec{\mathbf{k}}}}{\partial x}\right);$$

(B11)

is much more important because it links all three dimensions:

$$\left|\frac{\partial \delta E_z}{\partial y}\right| + \left|\frac{\partial \delta E_y}{\partial z}\right| \sim \left|\mathbf{k}_y \delta E_z\right| + \left|\mathbf{k}_z \delta E_y\right| \sim \frac{4\pi}{\vec{\mathbf{k}}^2} |\rho_{\vec{\mathbf{k}}}| \left(\frac{|\mathbf{k}_y|}{\gamma a_z} + \frac{|\mathbf{k}_z|}{a_y}\right);$$

$$\left|\frac{\partial \delta E_x}{\partial z}\right| + \left|\frac{\partial \delta E_z}{\partial x}\right| \sim \left|\mathbf{k}_z \delta E_x\right| + \left|\mathbf{k}_x \delta E_z\right| \sim \frac{4\pi}{\vec{\mathbf{k}}^2} |\rho_{\vec{\mathbf{k}}}| \left(\frac{|\mathbf{k}_x|}{\gamma a_z} + \frac{|\mathbf{k}_z|}{a_x}\right);$$

$$\left|\frac{\partial \delta E_x}{\partial y}\right| + \left|\frac{\partial \delta E_y}{\partial x}\right| \sim \left|\mathbf{k}_y \delta E_x\right| + \left|\mathbf{k}_x \delta E_y\right| \sim \frac{4\pi}{\vec{\mathbf{k}}^2} |\rho_{\vec{\mathbf{k}}}| \left(\frac{|\mathbf{k}_x|}{a_y} + \frac{|\mathbf{k}_y|}{a_x}\right).$$

(B12)

This allows us to estimate errors for each component of electric field:

$$|\vec{E}| \cong 4\pi \frac{|\rho_{\vec{\mathbf{k}}}|}{|\vec{\mathbf{k}}|}; \quad |\delta E_x| \sim |\vec{E}|\left(\frac{1}{|\vec{\mathbf{k}}|a_x} + \frac{1}{|\vec{\mathbf{k}}|a_y}\frac{|\mathbf{k}_y|}{|\mathbf{k}_x|} + \frac{1}{\gamma|\vec{\mathbf{k}}|a_z}\frac{|\mathbf{k}_z|}{|\mathbf{k}_x|}\right);$$

$$|\delta E_y| \sim |\vec{E}|\left(\frac{1}{|\vec{\mathbf{k}}|a_x}\frac{|\mathbf{k}_x|}{|\mathbf{k}_y|} + \frac{1}{|\vec{\mathbf{k}}|a_y} + \frac{1}{\gamma|\vec{\mathbf{k}}|a_z}\frac{|\mathbf{k}_z|}{|\mathbf{k}_y|}\right); |\delta E_z| \sim |\vec{E}|\left(\frac{1}{|\vec{\mathbf{k}}|a_x}\frac{|\mathbf{k}_x|}{|\mathbf{k}_z|} + \frac{1}{|\vec{\mathbf{k}}|a_y}\frac{|\mathbf{k}_y|}{|\mathbf{k}_z|} + \frac{1}{\gamma|\vec{\mathbf{k}}|a_z}\right);$$

(B13)

and

$$|\delta E_y| \sim |\vec{E}|\left(\frac{1}{\gamma|\vec{\mathbf{k}}|a_z}\frac{|\mathbf{k}_y|}{|\mathbf{k}_z|} + \frac{1}{|\vec{\mathbf{k}}|a_y}\right); |\delta E_y| \sim |\vec{E}|\left(\frac{1}{|\vec{\mathbf{k}}|a_y} + \frac{1}{|\vec{\mathbf{k}}|a_x}\frac{|\mathbf{k}_y|}{|\mathbf{k}_x|}\right);$$

$$|\delta E_z| \sim |\vec{E}|\left(\frac{1}{\gamma|\vec{\mathbf{k}}|a_z} + \frac{1}{|\vec{\mathbf{k}}|a_y}\frac{|\mathbf{k}_z|}{|\mathbf{k}_y|}\right); |\delta E_z| \sim |\vec{E}|\left(\frac{1}{\gamma|\vec{\mathbf{k}}|a_z} + \frac{1}{|\vec{\mathbf{k}}|a_x}\frac{|\mathbf{k}_z|}{|\mathbf{k}_x|}\right);$$

(B14)

$$|\delta E_x| \sim |\vec{E}|\left(\frac{1}{\gamma|\vec{\mathbf{k}}|a_z}\frac{|\mathbf{k}_x|}{|\mathbf{k}_z|} + \frac{1}{|\vec{\mathbf{k}}|a_x}\right); |\delta E_x| \sim |\vec{E}|\left(\frac{1}{|\vec{\mathbf{k}}|a_x} + \frac{1}{|\vec{\mathbf{k}}|a_y}\frac{|\mathbf{k}_x|}{|\mathbf{k}_y|}\right).$$

Now, let's introduce the following definitions:

$$\varepsilon_x = \frac{1}{|\vec{k}|a_x}; \varepsilon_y = \frac{1}{|\vec{k}|a_y}; \varepsilon_z = \frac{1}{\gamma|\vec{k}|a_z};$$

And

$$r_{xy} = \frac{|k_x|}{|k_y|}; r_{xz} = \frac{|k_x|}{|k_z|}; r_{yz} = \frac{|k_y|}{|k_z|};$$

(B15)

and rewrite (B13-14) as

$$|\delta E_x| \sim |\vec{E}| \cdot \left(\varepsilon_x + \frac{\varepsilon_y}{r_{xy}} + \frac{\varepsilon_z}{r_{xz}}\right); |\delta E_y| \sim |\vec{E}|\left(\varepsilon_x r_{xy} + \varepsilon_y + \frac{\varepsilon_z}{r_{yz}}\right); |\delta E_z| \sim |\vec{E}|(\varepsilon_x r_{xz} + \varepsilon_y r_{yz} + \varepsilon_z);$$

$$|\delta E_y| \sim |\vec{E}|(\varepsilon_z r_{yz} + \varepsilon_y); |\delta E_y| \sim |\vec{E}|\left(\varepsilon_y + \frac{\varepsilon_x}{r_{xy}}\right); |\delta E_z| \sim |\vec{E}|\left(\varepsilon_z + \frac{\varepsilon_y}{r_{yz}}\right); |\delta E_z| \sim |\vec{E}|\left(\varepsilon_z + \frac{\varepsilon_x}{r_{xz}}\right); \quad (B16)$$

$$|\delta E_x| \sim |\vec{E}|(\varepsilon_z r_{xz} + \varepsilon_x); |\delta E_x| \sim |\vec{E}|(\varepsilon_x + \varepsilon_y r_{xy}).$$

And finally, the combination of all estimations results in the following:

$$|\delta E_x| \sim |\vec{E}| \cdot \min\left(\varepsilon_x + \frac{\varepsilon_y}{r_{xy}} + \frac{\varepsilon_z}{r_{xz}}, \varepsilon_x + \varepsilon_z r_{xz}, \varepsilon_x + \varepsilon_y r_{xy}\right) \leq \varepsilon_x + \varepsilon_y + \varepsilon_z;$$

$$|\delta E_y| \sim |\vec{E}|\min\left(\varepsilon_x r_{xy} + \varepsilon_y + \frac{\varepsilon_z}{r_{yz}}, \varepsilon_y + \varepsilon_z r_{yz}, \frac{\varepsilon_x}{r_{xy}} + \varepsilon_y\right) \leq \varepsilon_x + \varepsilon_y + \varepsilon_z; \quad (B17)$$

$$|\delta E_z| \sim |\vec{E}|\min\left(\varepsilon_x r_{xz} + \varepsilon_y r_{yz} + \varepsilon_z, \frac{\varepsilon_y}{r_{yz}} + \varepsilon_z, \frac{\varepsilon_x}{r_{xz}} + \varepsilon_z\right) \leq \varepsilon_x + \varepsilon_y + \varepsilon_z;$$

where we took into account that $\min(r, r^{-1}) \leq 1, \forall r \geq 0$. It means that

$$|\vec{k}|a_x \gg 1; |\vec{k}|a_y \gg 1; \gamma_o|\vec{k}|a_z \gg 1;$$

$$a_{1,2} \cdot \sqrt{(\beta_o \gamma_o)^2 (k_1^2 + k_2^2) + k_3^2} \gg \beta_o \gamma_o; \quad a_3 \cdot \sqrt{(\beta_o \gamma_o)^2 (k_1^2 + k_2^2) + k_3^2} \gg 1.$$

(B18)

are sufficient conditions in the co-moving frame for Eq.(B7) to be a valid approximation for the electric field.

Lorentz transformation (B2) changes these conditions to the lab-frame as follows:

$$a_{x,y} \cdot \sqrt{\vec{k}_\perp^2 + \frac{\vec{k}_z^2}{\gamma^2}} \gg 1; \quad a_z \cdot \sqrt{\gamma^2 \vec{k}_\perp^2 + \vec{k}_z^2} \gg 1.$$

(B19)

These conditions are most important for the case of the longitudinal density modulation

$$\vec{k}_{lab} = \hat{z}k_{//}; \quad k_{//} \gg \max\left(\frac{\gamma}{a_{x,y}}, \frac{1}{a_z}\right),$$

(B20)

which means that transverse beam size can play important role in determining applicability of this important approximation.

Fig. A.1 provides intuitive illustrations of applicability and violations of conditions (B19).

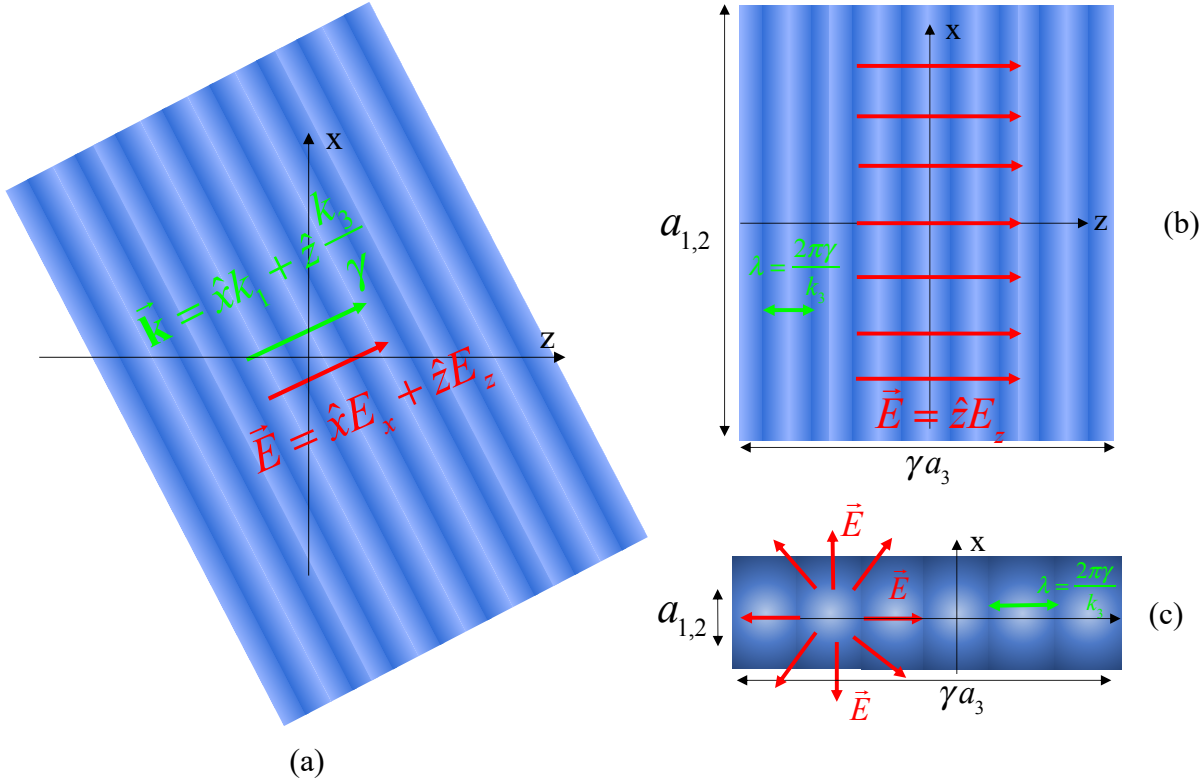

Figure A.1. (a) Geometrical explanation of proportionally between components of electric field, $\vec{E}$, and wave-vector, $\vec{k}$. Blue color represents a periodic density modulation in the direction of $\vec{k}$-vector. Figures on the right illustrate electric field structures (red arrows) in two cases: (b) when the condition (B19), $|a_{1,2} \cdot k_3| \gg \gamma_o$ is satisfied and electric field has plane-wave structure, and (c) when the condition (B19) is violated $a_{1,2} \sim \lambda = \dfrac{2\pi\gamma}{k_3}$, and the electric field is no longer parallel to vector k.

These findings have following foundation:
1. In the co-moving frame of the beam, where we are evaluating electric field, the length of the of the bunch is increased by factor $\gamma$, while value of longitudinal component of $k$-vector, $k_z$, is reduced by factor $\gamma$. In contrast, transverse components remain unchanged.
2. In a plane geometry of periodic density modulation, strength of components of electric field is directly proportional to the value of the corresponding component of k-vector in the comoving frame – see illustration in Fig. A.1. This is direct result solving electrostatic equation

$$div\vec{E} = 4\pi\tilde{\rho}_{\vec{k}} \cos\vec{k}\vec{r} \Rightarrow \vec{E} = \vec{k} \cdot \dfrac{4\pi\tilde{\rho}_{\vec{k}}}{|\vec{k}|^2} \sin\vec{k}\vec{r}.$$

3. It means that if component of the *k*-vector is zero, it is also true for the component of electric field. For example, longitudinal modulation with $\vec{k}_\perp = 0$, conditions (B19) are reduced to

$$|a_3 \cdot k_3| \gg 1; |a_{1,2} \cdot k_3| \gg \beta_o \gamma_o;$$

where fist condition is quite natural – and intuitive – requiring that there will be multiple oscillation in the longitudinal direction for Fourier components to stay uncoupled. The second condition is less obvious: it comes from the requirement that Fourier components of the electric field repeat the structure of Fourier components of density modulation, as can be seen in Figures A.1(a) and A.1(b). When this condition is violated, simple proportionality relations between density modulation and electric field brakes and it results in coupling of Fourier harmonics. Such coupling turns problem under consideration into an unsolvable.

While we illustrated importance of both conditions in Eq.(B19) for longitudinal modulation, the same considerations are valid for transverse directions.

**Appendix C. Expression for charge and current density modulation in laboratory frame**

Solving Maxwell equations require knowledge of the charge and current densities as functions of coordinates and time:

$$div\vec{B} = 0;\ div\vec{E} = 4\pi\rho;\ curl\vec{E} = -\frac{1}{c}\frac{\partial \vec{B}}{\partial t};\ curl\vec{B} = \frac{1}{c}\frac{\partial \vec{E}}{\partial t} + \frac{4\pi}{c}\vec{j};$$

$$\rho(\vec{r},t) = e\int f(\vec{r},\vec{v},t)d\vec{v}^3;\ \vec{j}(\vec{r},t) = e\int \vec{v}f(\vec{r},\vec{v},t)d\vec{v}^3, \quad (C1)$$

where $f(\vec{r},\vec{v},t)$ is the particles distribution function in the $(\vec{r},\vec{v})$ configuration space. Using $s$ as an independent variable makes the connection between $\rho, \vec{j}$ and the phase space distribution function $\tilde{f}(q,P,s)$ non-trivial, where $(q,P)$ is the conjugate Canonical set of coordinates and momenta. This Appendix is dedicated for establishing such connection and finding corresponding 4-potential of the EM field.

Let's introduce an instantaneous Cartesian coordinate system with the z-axis along the reference trajectory at $s = s_o$ (see Eq. (1) in the main text):

$$\hat{x} = \hat{e}_1(s_o);\ \hat{y} = \hat{e}_2(s_o);\ \hat{z} = \hat{e}_3(s_o) = \left.\frac{d\vec{r}_o(s)}{ds}\right|_{s=s_o};\ \frac{d\hat{e}_1}{ds} = K(s)\cdot\hat{e}_3 - \kappa(s)\cdot\hat{e}_2;\ \frac{d\hat{e}_2}{ds} = \kappa(s)\cdot\hat{e}_1;$$

$$\vec{r} = \vec{r}_o(s_o) + \hat{x}\cdot x + \hat{y}\cdot y + \hat{z}\cdot z \equiv \vec{r}_o(s) + \hat{e}_1 q_1 + \hat{e}_2(s)q_2; \quad (C2)$$

$$q_1 = \hat{e}_1(s)\cdot(\hat{x}\cdot x + \hat{y}\cdot y + \vec{r}_o(s_o) - \vec{r}_o(s));\ q_2 = \hat{e}_2(s)\cdot(\hat{x}x + \hat{y}\cdot y + \vec{r}_o(s_o) - \vec{r}_o(s));$$

$$q_3 = c(t_o(s) - t).$$

With the following ratios in the vicinity of $\vec{r}_o(s)$:

$$dx = dq_1 + q_2\kappa(s_o)ds;\ dy = dq_2 - q_1\kappa(s_o)ds;\ dz = (1 + q_1 K(s_o))ds;$$

$$q_3 = c(t_o(s) - t);\ dq_3 = \frac{ds}{\beta_o(s_o)} - cdt. \quad (C3)$$

where we used Eq. (1) for the reference trajectory. At fixed s:

$$ds = 0 \rightarrow dz = 0;\ dq_1 = dx;\ dq_1 = dy;\ dq_3 = -cdt. \quad (C4)$$

Number of particles confined in an infinitesimal volume $dq^3$ at fixed $s=s_o$ is defined as:

$$dn = |dq^3|\int \tilde{f}(q,P,s_o)dP^3 = cdxdydt\int \tilde{f}(q,P,s_o)dP^3 \quad (C5)$$

is identical to the number of particles passing though the elementary area $dxdy$ locates at $s=s_o$ in time interval $dt$:

$$\vec{j}(\vec{r},t) = \int \vec{v}f(\vec{r},\vec{v},t)d\vec{v}^3;\ dn = d\vec{a}\cdot\vec{j}\cdot dt;\ d\vec{a} = \hat{z}dxdy;$$

$$dn = dxdydt\int v_z f(\vec{r},\vec{v},t)d\vec{v}^3, \quad (C6)$$

resulting in

$$\int \tilde{f}(q,P,s)dP^3 = \frac{1}{c}\int v_z f(\vec{r},\vec{v},t)d\vec{v}^3 \tag{C7}$$

Using paraxial approximation $v_z = v_o + \delta v; |\delta v| \ll v_o$, and neglecting $\delta v$ in the integral (C7), we can express $\rho, \vec{j}$ using phases space distribution function as:

$$\rho(\vec{r},t) \cong \frac{\rho(q,s)}{\beta_o(s)}; \vec{j}(\vec{r},t) \cong \hat{z}c\rho(q,s); \rho(q,s) = e\int \tilde{f}(q,P,s)dP^3. \tag{C8}$$

Applying Fourier transformation[14]

$$g_{\vec{k}} \equiv g(k_1,k_2,k_3,s) = \int g(\vec{q},s)e^{-i\vec{k}\cdot\vec{q}}d\vec{q}^3$$

to (C8) we obtain expressions for $\rho, \vec{j}$ at fixed $s$:

$$\rho_{\vec{k}} = e\frac{\tilde{f}_{\vec{k}}}{\beta_o}; \vec{j}_{\vec{k}} = e\hat{z}c\tilde{f}_{\vec{k}}; \tilde{f}_{\vec{k}} \equiv \tilde{f}(\vec{k},s) = \int e^{-i\vec{k}\cdot\vec{q}}\tilde{f}(q,P,s)dP^3 dq^3;$$

$$\rho = \frac{1}{(2\pi)^3}\frac{1}{\beta_o}\int \tilde{f}_{\vec{k}}e^{i\vec{k}\cdot\vec{q}}d\vec{k}^3; \vec{j} = \frac{e\hat{z}c}{(2\pi)^3}\int \tilde{f}_{\vec{k}}e^{i\vec{k}\cdot\vec{q}}d\vec{k}^3. \tag{C9}$$

Let's calculate Fourier harmonic of the density

$$\rho(\vec{k},\omega) = \int \rho e^{-i(\vec{k}\cdot\vec{r}-\omega t)}d\vec{r}^3 dt = \frac{e}{(2\pi)^3}\int d\vec{k}^3 \int \frac{\tilde{f}_{\vec{k}}(s)}{\beta_o(s)}e^{i(\vec{k}\cdot\vec{q}-\vec{k}\cdot\vec{r}+\omega t)}d\vec{r}^3 dt. \tag{C10}$$

Using $q_1 = x; q_2 = y$ and combining terms in the exponent:

$$\vec{k}\cdot\vec{q} - \vec{\mathbf{k}}\cdot\vec{r} + \omega t = (k_1 - \mathbf{k}_x)x + (k_2 - \mathbf{k}_y)y + (\omega - k_3 c)t + k_3 ct_o(s) - \mathbf{k}_z z; \tag{C11}$$

makes expression integrable:

$$\frac{1}{(2\pi)^3}\iint e^{i(k_1-\mathbf{k}_x)x}e^{i(k_2-\mathbf{k}_y)y}e^{-i(\omega-ck_3)t}dx\,dy\,dt = \delta(k_1-\kappa_x)\delta(k_2-\kappa_y)\delta(\omega-ck_3);$$

$$\iint \delta(k_1-\mathbf{k}_x)\delta(k_2-\mathbf{k}_y)\delta(\omega-c\mathbf{k}_3)\tilde{f}(k_1,k_2,k_3,s)dk^3 = \frac{1}{c}\tilde{f}\left(\mathbf{k}_x,\mathbf{k}_y,\frac{\omega}{c},s\right); \tag{C12}$$

$$\rho(\vec{\mathbf{k}},\omega) = \frac{e}{c}\int \frac{\tilde{f}\left(\mathbf{k}_x,\mathbf{k}_y,\frac{\omega}{c},s\right)}{\beta_o(s)}e^{i(\omega t_o(s)-\mathbf{k}_z z)}dz.$$

---

[14] In this Appendix, we use interchangeably both the compact, $g_{\vec{k}}$, and detailed, $g(k_1,k_2,k_3,s)$, notation for the Fourier components defined in the accelerator coordinates.

At this point we can use our assumption that the scale of the variation of the accelerator parameters (such as trajectory curvature, $\beta_o$, etc.) are much larger than that of the modulation. In addition, we assume that evolution of the density modulation as function of $s$ is also much slower than fast oscillating term $e^{i\omega t_o(s)}$. This assumption will allow us to move $\tilde{f}$ and $\beta_o$ outside the integral and also to expand the arrival time of the reference particle with respect to the azimuth $\bar{s}$, where we locate the origin of z-axis:

$$t_o(s) \cong t_o(\bar{s}) + \frac{z}{c\beta_o(\bar{s})}, \tag{C13}$$

and arrive to the final relation between Fourier components in two systems of coordinates:

$$\rho(\vec{k},\omega) = \frac{2\pi e}{c}\tilde{f}(\mathbf{k}_x,\mathbf{k}_y,\beta_o\mathbf{k}_z,\bar{s})\delta(k_o - \beta_o\mathbf{k}_z)e^{i\beta_o\mathbf{k}_z ct_o(\bar{s})}; \quad k_o = \frac{\omega}{c}; \tag{C14}$$

$$\vec{j}(\vec{k},\omega) = \hat{z}c\beta_o\rho(\vec{k},\omega)$$

where we used $\delta\left(\frac{k_o}{\beta_o} - \mathbf{k}_z\right) = \beta_o\delta(k_o - \beta_o\mathbf{k}_z)$ and singularity of Dirac's $\delta$-function:

$$g(x)\delta(x-y) = g(y)\delta(x-y).$$

To find 4-potential induced by such perturbation we can use Lorenz gauge $\frac{\partial\varphi}{c\partial t} + div\vec{A} = 0$ providing for separation of equations for each component of 4-potentail [37] [15]:

$$\frac{\partial^2\varphi}{c^2\partial t^2} - \Delta\varphi = 4\pi\rho; \quad \frac{\partial^2\vec{A}}{c^2\partial t^2} - \Delta\vec{A} = 4\pi\vec{j},$$

which can be Fourier transformed to:

$$\varphi(\vec{k},\omega)e^{i(\vec{k}\vec{r}-\omega t)} = \frac{4\pi\rho(\vec{k},\omega)}{\vec{k}^2 - k_o^2}e^{i(\vec{k}\vec{r}-\omega t)} = \frac{8\pi^2 e}{c}\frac{\tilde{f}(\mathbf{k}_x,\mathbf{k}_y,\beta_o\mathbf{k}_z,\bar{s})}{\vec{k}^2 - \beta_o^2\mathbf{k}_z^2}\delta(k_o - \beta_o\mathbf{k}_z)e^{i(\vec{k}\vec{r}+\omega(t_o(\bar{s})-t))}; \tag{C15}$$

$$\vec{A}(\vec{k},\omega) = \vec{z}\varphi(\vec{k},\omega)\frac{k_o}{\mathbf{k}_z} = \vec{z}\beta_o\varphi(\vec{k},\omega).$$

In the inverse Fourier transform

$$\varphi(\vec{r},t) = \frac{1}{(2\pi)^4}\int\varphi(\vec{k},\omega)e^{i(\vec{k}\vec{r}-\omega t)}d\omega d\mathbf{k}^3,$$

$\delta$-function makes integral over $\omega$ straight forward:

---

[15] The Lorentz gauge can be used for time-dependent component of the EM field, which is of interest in this paper.

$$\frac{1}{2\pi}\int e^{i\omega(t_o(\bar{s})-t)}g\left(\vec{\mathbf{k}},\frac{\omega}{c}\right)\delta\left(\frac{\omega}{c}-\beta_o\mathbf{k}_z\right)d\omega=\frac{c}{2\pi}g(\vec{\mathbf{k}},\beta_o\mathbf{k}_z)e^{i\beta_o\mathbf{k}_z\tau(\bar{s})};\tau(\bar{s})=c(t_o(\bar{s})-t), \quad (C16)$$

with the remaining integral of

$$\varphi(\vec{r},t)=4\pi e\int\frac{\tilde{f}(\mathbf{k}_x,\mathbf{k}_y,\beta_o\mathbf{k}_z,\bar{s})}{\vec{\mathbf{k}}^2-\beta_o^2\mathbf{k}_z^2}e^{i(\vec{\mathbf{k}}\vec{r}+\beta_o\mathbf{k}_z\tau(\bar{s}))}\frac{d\mathbf{k}^3}{(2\pi)^3};\ \vec{A}(\vec{r},t)=\vec{z}\beta_o\varphi(\vec{r},t). \quad (C17)$$

Taking into account expansion (C13), the exponent in (C17) can be expressed using the accelerator coordinates:

$$\vec{\mathbf{k}}\cdot\vec{r}+\beta_o\mathbf{k}_z\tau(\bar{s})=\vec{k}\cdot\vec{q};k_{1,2}=\mathbf{k}_{x,y};k_3=\beta_o\mathbf{k}_z; \quad (C18)$$

and using ratio $\beta_o d\mathbf{k}^3=dk^3$ we get expression connecting the 4-potentail and density perturbation in the accelerator coordinates:

$$\varphi(\vec{q},s)\equiv\varphi(\vec{r},t)=4\pi e\beta_o\gamma_o^2\int\frac{\tilde{f}_{\vec{k}}}{\gamma_o^2\beta_o^2\vec{k}_\perp^2+k_3^2}e^{i\vec{k}\cdot\vec{q}}\frac{d\vec{k}^3}{(2\pi)^3};\ \vec{A}(\vec{q},s)=\vec{z}\beta_o\varphi(\vec{q},s). \quad (C19)$$

**Appendix D. Perturbed Hamiltonian**

As derived in Appendix C, density perturbation results in an additional 4-potentail

$$\delta\varphi^i=\{\delta\varphi,\delta\vec{A}\};\delta\vec{A}=\hat{z}\beta_o\delta\varphi. \quad (D1)$$

which we will consider being infinitesimally small: $\delta\varphi\sim O(\varepsilon),\varepsilon\ll 1$. The goal of this Appendix is to define an additional term of the reduced accelerator Hamiltonian (6) resulting from the density perturbation:

$$\begin{aligned}h^*=&-(1+Kq_1)\sqrt{\frac{(E_o+cP_3-e\varphi_\perp-e\delta\varphi)^2}{c^2}-m^2c^2-\left(P_1-\frac{e}{c}A_1\right)^2-\left(P_2-\frac{e}{c}A_2\right)^2}\\&-\frac{e}{c}(1+Kq_1)(A_z+\beta_o\delta\varphi)+\kappa q_1P_2-\kappa q_2P_1-\frac{c}{v_o(s)}P_3+q_3\frac{d}{ds}\left(E_o(s)+e\frac{\varphi(\vec{r}_o(s),t)}{c}\right)\end{aligned} \quad (D2)$$

where we used the explicit expression for $A_3$ component of the vector potential (7). Perturbation of the Hamiltonian is coming only from the first two terms in r.h.s of (D2).

$$\delta h^* = -(1+Kq_1)\left(\sqrt{\frac{(E-e\delta\varphi)^2}{c^2} - m^2c^2 - \vec{p}_\perp^{\,2}} - p_z + \frac{e}{c}\beta_o\delta\varphi\right); \quad p_z = \sqrt{\frac{E^2}{c^2} - m^2c^2 - \vec{p}_\perp^{\,2}};$$

$$\sqrt{\frac{(E-e\delta\varphi)^2}{c^2} - m^2c^2 - \vec{p}_\perp^{\,2}} - p_z = -\frac{E}{cp_z}\frac{e}{c}\delta\varphi + O(\delta\varphi)^2; \tag{D3}$$

$$E = H - e\varphi;\ \beta_o = \frac{v_o}{c};\ \beta_z = \frac{v_z}{c} = \frac{cp_z}{E};\ 1-\beta_o^2 = \gamma_o^{-2};$$

$$\delta h^* = (1+Kq_1)\frac{e}{c}\delta\varphi\left(\frac{c}{v_z} - \frac{v_o}{c}\right) = \frac{1}{\beta_o\gamma_o^2}\frac{e}{c}\delta\varphi\left\{1+\gamma_o^2\left(\frac{\beta_o}{\beta_z}-1\right)\right\}(1+Kq_1).$$

First, in paraxial approximation term $|Kq_1| \ll 1$ can be dropped. It is also easy to show that the second term in the curly brackets is infinitesimally small in the case of paraxial motion resulting in non-relativistic motion in the co-moving frame:

$$\gamma_o^2\left(\frac{\beta_o}{\beta_z}-1\right) = \frac{\gamma_o^2\vec{\beta}_\perp^{\,2}}{\beta_z(\beta_o+\beta_z)} + \frac{\gamma-\gamma_o}{\gamma\beta_z(\beta_o+\beta_z)} \sim \frac{\vec{\beta}_{cm\perp}^{\,2}}{2} + \frac{\delta\gamma}{2\gamma} \ll 1. \tag{D4}$$

Specifically, $c\gamma_o\vec{\beta}_\perp$ is the transverse velocity in the co-moving frame and $\frac{\delta\gamma}{\gamma}$ is the relative energy deviation in the beam. Both of these values are assumed to be infinitesimally small. As the result, the perturbation of the Hamiltonian is reduced to:

$$\delta h^* = \frac{1}{\beta_o\gamma_o^2}\frac{e}{c}\delta\varphi = \frac{4\pi e^2}{c}\int\frac{\tilde{\rho}_{\vec{k}}}{\gamma_o^2\beta_o^2\vec{k}_\perp^{\,2} + k_z^2}e^{i\vec{k}\vec{q}}\frac{d\vec{k}^3}{(2\pi)^3}; \tag{D5}$$

$$\tilde{\rho}_{\vec{k}} = \int e^{-i\vec{k}\vec{q}}\tilde{f}(q,p,s)dp^3dq^3.$$

**Appendix E. Numerical solution for linear integral equation**

While this is known that linear integral equations are relatively easy to solve, for completeness we describe a simple, by design, step by step process for our specific case. More sophisticated methods can be found in ref. [43-44].

Let's split our accelerator in small segments $\Delta s$. We start from s=0 and evaluate $\tilde{\rho}_{\vec{k}_o}(0)$, assuming a known infinitesimal perturbation of initial beam density $\tilde{f}_o(q,P)$:

$$\tilde{\rho}_o = \tilde{\rho}_{\vec{k}}(0) = \int e^{i\vec{k}\vec{q}}dq^3\,dP^3\tilde{f}_o(q,P) \equiv \int e^{i\vec{k}\vec{q}}dq^3\,dP^3\tilde{f}_o(q,P). \tag{E1}$$

Next, we will use transport matrices at $s_1 = \Delta s$ to evaluate all other evolving components: step i=1

$$\mathbf{A}_0 = \mathbf{D}_0 = \mathbf{I}; \mathbf{B}_0 = \mathbf{C}_0 = 0; \mathbf{A}_1 = \mathbf{A}(\Delta s); \mathbf{B}_1 = \mathbf{B}(\Delta s); \mathbf{C}_1 = \mathbf{C}(\Delta s); \mathbf{D}_1 = \mathbf{D}(\Delta s);$$
$$\vec{k}_1 = \vec{k}(\Delta s) = \underline{\vec{k}} \cdot \mathbf{A}^{-1}(\Delta s); \vec{k}_o \equiv \underline{\vec{k}}$$
$$\tilde{\rho}_{\vec{k}_o}(\Delta s) = \frac{1}{\det \mathbf{A}_1} \int e^{i\vec{k}(\Delta s)\vec{q}} \, dq^3 \, dP^3 \underline{\tilde{f}}_o \left( \mathbf{D}_1^T q - \mathbf{B}_1^T p, -\mathbf{C}_1^T q + \mathbf{A}_1^T p \right);$$
$$\upsilon_1 = \gamma_o(\Delta s)^2 \beta_o(\Delta s)^2 \vec{k}_\perp(\Delta s)^2 + k_3(\Delta s)^2; \quad u_1 = \vec{k}_1 \cdot \vec{\mathbf{B}}_1 \cdot \underline{\vec{k}}; u_o = u(0) = 0.$$
(E2)

For known background initial momentum distribution, one can calculate Landau damping term

$$\vec{\eta}_1 = \vec{k}_1 \cdot \vec{\mathbf{B}}_1; \vec{\eta}_0 = 0; L_{10} = \int e^{i(\vec{\eta}_1 - \vec{\eta}_o) \cdot \vec{P}} \underline{f}_o(P) dP^3 \tag{E3}$$

and calculate density modulation at step 1:

$$\tilde{\rho}_1 = \tilde{\rho}(\Delta s, \vec{k}(\Delta s)) = -\frac{4\pi n_o e^2}{c \det \mathbf{A}_0 \upsilon_0} L_{10}(u_1 - u_0) \Delta s \cdot \tilde{\rho}_o + \tilde{\rho}_{\vec{k}_o}(\Delta s). \tag{E4}$$

Let's assume that we competed step $i=n-1$ and are going to next step $s_n = n\Delta s$:
$$\mathbf{A}_n = \mathbf{A}(n\Delta s); \mathbf{B}_n = \mathbf{B}(n\Delta s); \mathbf{C}_n = \mathbf{C}(n\Delta s); \mathbf{D}_n = \mathbf{D}(n\Delta s); \vec{k}_n = \vec{k}(n\Delta s) = \vec{k}_o \cdot \mathbf{A}^{-1}(n\Delta s);$$
$$\tilde{\rho}_{\vec{k}_o}(s_n) = \frac{1}{\det \mathbf{A}_1} \int e^{i\vec{k}(\Delta s)\vec{q}} \, dq^3 \, dp^3 \underline{f}_o \left( \mathbf{D}_n^T q - \mathbf{B}_n^T p, -\mathbf{C}_n^T q + \mathbf{A}_n^T p \right);$$
$$\upsilon_n = \gamma_o(n\Delta s)^2 \beta_o(n\Delta s)^2 \vec{k}_{n\perp}^2 + k_{n3}^2;$$
$$u_n = \vec{k}_n \cdot \vec{\mathbf{B}}_n \cdot \vec{k}_o;$$
(E5)

and calculate all relevant Landau damping terms for propagation from $s_i$ to $s_n$;

$$\vec{\eta}_n = \vec{k}_n \cdot \vec{\mathbf{B}}_n; L_{ni} = \int e^{i(\vec{\eta}_n - \vec{\eta}_i) \cdot \vec{P}} \underline{f}_o(P) dP^3 \tag{E6}$$

and calculate density perturbation as a sum

$$\tilde{\rho}_n = \tilde{\rho}(s_n, \vec{k}_n) = -\frac{4\pi n_o e^2 \Delta s}{c} \sum_{i=0}^{n-1} \frac{L_{ni}(u_n - u_i) \cdot \tilde{\rho}_i}{\det \mathbf{A}_i \upsilon_i} + \tilde{\rho}_{\vec{k}_o}(s_n). \tag{E7}$$

This process is iterative with all information about density evaluation prepared at previous steps. What is shown here is a process with first order of precision, but it can be improved using higher order procedures.